\documentclass[sigconf]{acmart}

\usepackage{acmart-taps}
\usepackage{graphics}
\usepackage{xspace}
\usepackage{enumitem}
\usepackage{subcaption}
\usepackage{color, soul}
\usepackage{bm}

\def\S{Section\xspace}

\def\insitu{\textit{in situ}\xspace}
\def\ie{\textit{i.e.,}\xspace}
\def\etal{\textit{et~al.}\xspace}

\def\eg{\textit{e.g.,}\xspace}
\def\incl{\textit{incl.}\xspace}

\def\insitu{\textit{in-situ}\xspace}

\newcommand{\red}[1]{\textcolor{black}{#1}}

\colorlet{agent}{blue!50}
\colorlet{participant}{gray}

\colorlet{mr}{orange!25}
\colorlet{mr-tp1}{green!25}
\colorlet{mr-tp2}{blue!25}

\begin{document}
\copyrightyear{2026}
\acmYear{2026}
\setcopyright{cc}
\setcctype{by}
\acmConference[UIST '26]{The 39th Annual ACM Symposium on User Interface Software and Technology}{November 02--05, 2026}{Detroit, MI, USA}
\acmBooktitle{The 39th Annual ACM Symposium on User Interface Software and Technology (UIST '26), November 02--05, 2026, Detroit, MI, USA}
\acmDOI{10.1145/3830398.3830487}
\acmISBN{979-8-4007-2856-3/2026/11}



\author{Shaoze Zhou}
\orcid{0009-0000-3243-0599}
\email{szhou010@fiu.edu}
\affiliation{%
  \institution{Florida International University}
  \city{Miami}
  \state{FL}
  \country{USA}
}

\author{Joaquin Frangi}
\orcid{0009-0002-5923-3573}
\email{jfran348@fiu.edu}
\affiliation{%
  \institution{Florida International University}
  \city{Miami}
  \state{FL}
  \country{USA}
}

\author{Diana Nelly Rivera Rodriguez}
\orcid{0009-0000-5369-6863}
\email{drive217@fiu.edu}
\affiliation{%
  \institution{Florida International University}
  \city{Miami}
  \state{FL}
  \country{USA}
}

\author{Rawan Alghofaili}
\orcid{0000-0001-6510-4562}
\email{rawan@utdallas.edu}
\affiliation{%
  \institution{University of Texas at Dallas}
  \city{Richardson}
  \state{TX}
  \country{USA}
}

\author{Janet G. Johnson}
\orcid{0000-0002-0456-4028}
\email{jgjanet@umich.edu}
\affiliation{%
  \institution{University of Michigan}
  \city{Ann Arbor}
  \state{MI}
  \country{USA}
}

\author{Lingyao Li}
\orcid{0000-0001-5888-8311}
\email{lingyaoli@arizona.edu}
\affiliation{%
  \institution{University of Arizona}
  \city{Tucson}
  \state{AZ}
  \country{USA}
}

\author{Renkai Ma}
\orcid{0000-0002-4434-2235}
\email{mark@ucmail.uc.edu}
\affiliation{%
  \institution{University of Cincinnati}
  \city{Cincinnati}
  \state{OH}
  \country{USA}
}

\author{Christine Lisetti}
\orcid{0000-0003-0756-133X}
\email{lisetti@fiu.edu}
\affiliation{%
  \institution{Florida International University}
  \city{Miami}
  \state{FL}
  \country{USA}
}

\author{Chen Chen}
\orcid{0000-0001-7179-0861}
\email{chechen@fiu.edu}
\affiliation{%
  \institution{Florida International University}
  \city{Miami}
  \state{FL}
  \country{USA}
}

\renewcommand{\shortauthors}{Zhou~\etal}

\begin{teaserfigure}
    \centering
    \includegraphics[width=\linewidth]{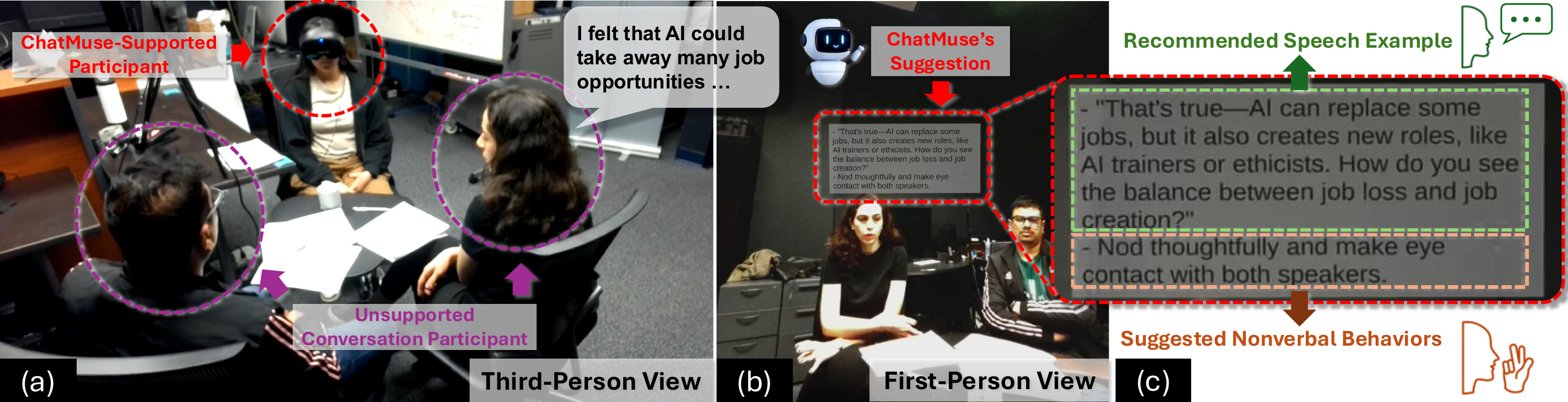}
    \caption{\sysname~provides real-time MR information support for in-person small-group conversations; (a) a third-person view of a three-participant conversation; (b) a first-person view from the MR user; (c) example supporting information.}
    \label{fig::teaser}
\end{teaserfigure}

\def\sysname{ChatMuse}
\title[\sysname]{\sysname: Supporting In-Person Small-Group Conversation Experience with a Proactive Assistive AI Agent in Mixed Reality}
\soulregister\sysname0

\begin{abstract}

In-person small-group conversations play a crucial role in social interaction.
However, achieving effective group conversations can be challenging and cognitively demanding.
While \textbf{M}ixed \textbf{R}eality~(MR) headsets show promise as a conversational support system by presenting relevant information through MR overlays, it remains unclear how such supporting information should be designed for in-person \emph{group} conversations.
We propose \sysname, an MR-based proactive assistive system for in-person small-group conversations.
\sysname~analyzes verbal and nonverbal cues from all conversation participants and proactively provides guides on the user's verbal and nonverbal behaviors.
The behavioral responses of the supported users are then used to improve \sysname's support capabilities in subsequent interactions.
Our within-subject studies alongside focus groups with six participant groups ($N = 18$) revealed seven key benefits that \sysname~provides in supporting users' engagement and contribution to in-person small-group conversations.
Our research around \sysname~represents a design exploration of a new interaction space that investigates the feasibility of supporting in-person small-group conversations through a proactive assistive AI agent in MR.
\end{abstract}

\settopmatter{printfolios=false} 
\maketitle

\section{Introduction}

\begin{quote}
{\it ``Talking to people --- truly talking --- is a skill. Group conversations are like a game of jump rope; you have to find the right rhythm to jump in, or you just end up getting hit in the face.''} \hfill--- Amy Meredith Poehler
\end{quote}

\vspace{4px}

In-person small-group conversations in everyday and professional settings are critical to share information, spark new ideas, and build interpersonal relationships~\cite{Daft1986, Hasson2012, Sacks1974, Singh2025, Kraut1990}.
However, achieving effective in-person small-group conversations can be challenging and cognitively demanding, as the real-time nature requires full attention, provides no editing time, and involves understanding of nonverbal cues~\cite{Yi2021, Cooney2020, Daniel2006}. 
As a result, conversation participants need to understand complex interaction dynamics, adhere to social norms, manage multiple speakers, and tailor spoken messages for a broader audience~\cite{groupconversationchallenge}.

Recent \textbf{M}ixed \textbf{R}eality (MR) headsets show promise as conversational supporting systems by presenting relevant information through MR overlays.
In dyadic conversation, prior research has examined how critical communication support, such as keywords extracted from spoken speech~\cite{Fujimoto2025ChatAR, Jadon2024RealityChat}, relevant background context~\cite{Jadon2024RealityChat}, and guidance for subsequent utterances~\cite{Zhang2025Understood, Yang2025SocialMind}, can be used to facilitate in-person conversation.
However, it remains unclear how a similar conversational support can be designed for in-person \emph{group} conversations. 
Generalizing the design of MR-based conversational support systems from dyadic to group conversations is challenging: 
compared to dyadic conversations, in-person group conversations involve more complex dynamics due to the well-known ``many minds problem,'' in which communication and interaction challenges increase as more participants join, making turn-taking, listening, and disclosure more difficult~\cite{Cooney2020, Sun2020}.

Designing proactive support for a variety of conversational experiences is promising to enhance synchronous online~\cite{Liu2025, Houde2025, Leong2024} and in-person~\cite{Yang2025SocialMind, Rayan2025CueingCrowd} dyadic and group interactions.
\emph{Proactive AI}, rooted in the framework of mixed-initiative interaction~\cite{Allen1999MixedInitiativeInteraction, Horvitz1999, Novick1997}, envisions an AI agent that can autonomously determine when and how to act without requiring explicit user requests.
Recent conversational AI agents such as GPT-Live~\cite{gpt-Live} have demonstrated the capability to provide proactive backchanneling.
While prior work, \eg~SocialMind~\cite{Yang2025SocialMind}, has demonstrated the design of proactive conversational support systems for \emph{dyadic} interactions --- where real-time visual suggestions are rendered on heads-up displays --- it remains unclear how such designs can be generalized to \emph{group} conversations;
unlike dyadic settings, group conversations involve more than two participants and introduce more complex interaction dynamic; designing support for group conversations therefore requires addressing additional challenges beyond those encountered in dyadic interactions, such as the need to address the long-standing many minds problem~\cite{Cooney2020}, to adapt to the behaviors of multiple participants and conversational partners, and to manage issues like participation inequality~\cite{Cooney2020, Haan2021}.

We propose \emph{\sysname}\footnote{Linguistically, \emph{muse} often refers to the personified force who is the source of inspiration~\cite{musedefinition}. We use Chat\underline{Muse} to metaphorically describe how a proactive AI agent can intervene in and facilitate in-person small-group conversations.}, a novel MR-based proactive supporting system for in-person small-group conversations.
Figure~\ref{fig::teaser} illustrates an example experience in which \sysname~ provides suggestions to an MR user during a three-person group conversation on the impact of AI on the future of work.
During a conversation, \sysname~analyzes both verbal and nonverbal cues from all participants and proactively provides real-time guidance on speech and nonverbal behaviors for the MR user.
User's verbal and behavioral responses are then used to improve \sysname's support capabilities in subsequent interactions.

Grounded in user-centered~\cite{Norman1996UCSD} and participatory design~\cite{Muller1993} processes, we first explored key \textbf{D}esign \textbf{C}onsiderations (DCs) by conducting a formative study ($N = 10$) using the provotype \cite{Boer2012} and focus group~\cite{Krueger2014}.
We developed two testbeds using Meta Quest 3S~\cite{metaquest3s} to investigate the affordances of proactive information support and the types of information users might seek while interacting with multiple conversation participants.
Three DCs were identified concerning how to maximize the benefits of proactive information support and how to effectively design supporting information.

We then prototyped \sysname~on Meta Quest Pro~\cite{metaquestpro, metaquestpronews} and conducted within-subject studies alongside focus groups with six participant groups ($N = 18$) to evaluate the utility and usability of the AI-generated conversational support rendered as an MR overlay. 
By analyzing the focus group discussions alongside the support information and the rationales inferred by the AI agent, we identified seven key benefits that \sysname~provided to support users' engagement and in-person small-group conversations.
While most participants found \sysname~helpful, both in the content of the support information and how the agent determined when to deliver it, some noted instances where support was provided unnecessarily or, conversely, where needed support was not delivered.

Our research around \sysname~represents a design exploration of a new interaction space that investigates the feasibility of supporting in-person small-group conversations through a proactive assistive AI agent in MR.
Overall, we contribute:
\textbf{\emph{(i)}}~\textbf{an interactive experience} that supports users in engaging and participating in small-group conversations through private AI-driven information support;
\textbf{\emph{(ii)}}~\textbf{a pipeline design} that represents key group conversation contexts for an LLM agent, allowing \sysname~to capture and interpret both verbal and nonverbal behaviors by leveraging the reasoning and social behaviors of LLMs;
\textbf{\emph{(iii)}}~\textbf{a mixed-methods understanding} of how and to what extent AI can leverage complex contexts to support group conversants.

\section{Related Works}\label{sec::related}

\subsection{The Challenge of In-Person Conversation}\label{sec::related::group_conversation}

An effective group conversation should be purposeful and focused, and involve active participation, along with respectful communication, such as active listening and constructive feedback~\cite{Teamskill2025}.
Engaging effectively in group conversations can be challenging and cognitively demanding~\cite{Daniel2006, Bales1950, Elsayed1997}.
First, conversation extends beyond spoken words. Mehrabian's rule~\cite{Mehrabian1971, Mehrabian1967} highlights the importance of nonverbal signaling in interpersonal communication, such as vocal tone and body language.
Adding more participants to a conversation also creates challenges for both speakers and listeners, a phenomenon often referred to as the many minds problem~\cite{Cooney2020}. 
Compared with dyadic conversations, larger groups reduce the airtime available to each participant, make turn-taking increasingly difficult, and weaken or obscure backchannel feedback.
Second, individuals may misinterpret the context of a conversation due to differences in background knowledge or misunderstandings of nonverbal cues.
Those experiencing social withdrawal may feel uncomfortable --- or even anxious --- when engaging in conversations~\cite{Rubin2001}, which can hinder their ability to build relationships in daily life and collaborate effectively in professional settings.

\subsection{Supporting Group Conversation with MR}\label{sec::related::mr_conversation}
Prior work has explored the use of MR to support in-person conversations in both dyadic and group settings.

\vspace{2px}\noindent{\bf Dyadic conversation.}
For example, in a dyadic interaction, ChatAR \cite{Fujimoto2025ChatAR} rendered supplementary information derived from speech keywords within a display region positioned relative to the facial location of the conversational partner.
RealityChat~\cite{Jadon2024RealityChat} introduced a HoloLens-based system that visualized speech keywords and displayed supplementary images when specific keywords were selected.
SocialMind~\cite{Yang2025SocialMind} demonstrated how an LLM-based virtual assistant could provide \insitu~social assistance during live dyadic interactions.
Empathetic AuRea~\cite{Valente2022EmpathocAurea} employed colored ring overlays to visualize and amplify the emotions of the conversational partner. 
Understood~\cite{Zhang2025Understood} introduced a HoloLens-based system designed to assist individuals with ADHD during conversations by providing real-time summarization, context-aware word suggestions, and reminders for topic shifts.
WSCoach~\cite{Zhang2025WSCoach} used real-time auditory feedback delivered via a wearable device to help users reduce unwanted words in everyday communication.
A second line of research explored how MR notifications can be designed to support face-to-face dyadic interactions.
Janaka~\etal~\cite{Janaka2022} investigated MR notification placement, while ParaGlassMenu~\cite{Cai2023} introduced a circular menu overlay around the partner, both aiming to enable MR interaction while minimizing disruptions to ongoing conversations.
Despite the validated effectiveness in dyadic interactions, it remains challenging to generalize these designs to \emph{group} settings, where users must attend to multiple conversational partners.

\vspace{2px}\noindent{\bf Group conversation.} 
Researchers have explored group conversations with more than two participants.
Zhou~\etal~\cite{Zhou2026ChatMuseNeed} demonstrated the need for and value of providing private information support to facilitate in-person group conversations.
Logue~\etal~\cite{Damian2015Logue} employed Google Glass to provide presenters with real-time behavioral feedback in scenarios such as public speaking.
ConversAR~\cite{Bendarkawi2025ConversAR} presented an AR system to support group conversations for second language learners, providing real-time translations of their partners' speech.
Commercial MR applications, \eg~Spatial~\cite{spatial}, have explored shared workspaces to facilitate group conversations for the future of work.
Johnson~\etal~\cite{Johnson2025} examined the potential of collaborative humanoid AI agents to facilitate group conversations.
Building on a similar in-person small-group conversation context, but not as a shared facilitator, \sysname~examines the feasibility of providing real-time private conversational textual support to the user, controlled by a proactive AI agent.

\subsection{Supporting Group Conversation with AI}\label{sec::related::ai_conversation}

While the studies above focus on in-person group conversations, a complementary line of research has investigated the use of AI to enhance remote conversations, typically mediated through videoconferencing and \textbf{I}nstant \textbf{M}essaging~(IM).
Application design in these domains provides insights that are transferable to \sysname.

\vspace{2px}\noindent{\bf Multi-party videoconferencing.}
Existing research has demonstrated how AI agents can be integrated to understand the social role of conversational participants~\cite{Vinciarelli2011}, provide on-demand highlights, summaries, and actions~\cite{TeamsCoPilot, ZoomAICompanion, Moran1997, Kalnikaitundefined2012, Asthana2025}, reflect goals~\cite{Chen2025}, track social dynamics~\cite{Bhattacharya2018}, as well as address issues of inclusion and participation~\cite{Fu2022, Guo2019, Kohl2023, Muller2018}.
Others have considered critical nonverbal cues, including emotions~\cite{Zhou2022}, bodily gestures~\cite{Choi2021}, and environmental factors~\cite{Constantinides2020}.
Commercial applications such as Teams and Zoom have introduced reactive AI agents --- Copilot~\cite{TeamsCoPilot} and AI Companion~\cite{ZoomAICompanion} --- that allow users to request information during group meetings. 
More recently, CoExplorer~\cite{Park2024CoExplorer} introduced a GenAI-powered adaptive interface that optimized task space based on users' goal-oriented input and real-time activity context.

\vspace{2px}\noindent{\bf Instant Messaging~(IM).}
AI agents have also been explored to facilitate group conversations in IM applications. These AI agents often exhibit context-aware behaviors that support group conversation, through integrated memory, planning, and reflection~\cite{Park2023GenAgent, Wei2023MultiPartyChat}.
\mbox{Hohenstein}~\etal~\cite{Hohenstein2018} investigated AI-supported messaging through Google Allo's ``smart reply'' feature, which generated suggested responses by analyzing the conversation history.
Kim~\etal~\cite{Kim2020} designed a facilitator chatbot to enhance group discussions by structuring discussion time, promoting equitable participation, and organizing participants' ideas.
Others, \eg~\cite{Lee2025, Lee2025criticalthinking}, showed that AI chatbots could promote more equitable participation in group decision-making and support well-being-oriented peer interactions~\cite{Liu2024ComPeer}.
A few recent works have explored the feasibility of AI agents in \emph{proactive} roles, building on earlier research in mixed-initiative interactions that envisioned agents autonomously determining when and how to act~\cite{Allen1999MixedInitiativeInteraction, Horvitz1999, Liao2023}.
In the context of group conversations, Houde~\etal~\cite{Houde2025} developed a conversational agent in the form of a Slack bot to investigate how proactive and reactive agent interventions influence group ideation.
Liu~\etal~\cite{Liu2025} investigated a proactive agent capable of generating inner thoughts by interpreting the conversational context.
In contrast to the prior research, we focus on MR-mediated, in-person small-group conversations.
Designing support for \emph{in-person} conversations differs from text-based communication, as it requires addressing additional challenges arising from rich nonverbal contexts.
Beyond spoken speech, designing \sysname~requires an understanding of nonverbal cues.
\section{Formative Study}\label{sec::formative}

Grounded in user-centered design ~\cite{Norman1996UCSD}, we first explored the key DCs of \sysname~by addressing two \textbf{R}esearch \textbf{Q}uestions (RQs): {\it how should proactive support be designed, and how can it facilitate in-person group conversation experiences?} \textbf{(RQ1)}; and {\it what types of information are needed to support in-person group conversation experiences?} \textbf{(RQ2)}

\subsection{Study Design}\label{sec::formative::methods}

We recruited $10$~participants (age, $M = 25.5$, $SD = 3.5$) and formed two small groups with three participants each and one larger group with four participants (Figure~\ref{tab::formative-participants}, Appendix~\ref{app::participants}).
FP\# and FG\# were used to indicate \textbf{F}ormative study \textbf{P}articipants and participant \textbf{G}roup.
Participants were first introduced to the project.
Each group was then instructed to complete three phases.

\vspace{2px}
\noindent\textbf{Phase 1: Onboarding.}
Participants were asked to complete demographic questionnaires and the \textbf{I}nterpersonal \textbf{C}ommunication \textbf{S}kills \textbf{I}nventory (ICSI) --- a standard questionnaire that assesses individuals' communication skills~\cite{Interpersonal-Communication-Skills-Inventory, Bienvenu1971, Boyd1977} (Appendix~\ref{app::questionaires}).
Understanding participants' communication skills allowed us to better interpret the qualitative data.

\vspace{2px}\noindent{\bf Phase 2: Focus Group.}
Focus groups were conducted with each participant group.
We focused on two guiding questions:
\textbf{(1)} {\it Given your prior experience, when do you think group conversation dynamics usually deteriorate, and what is your strategy to enhance the conversation experience?}
\textbf{(2)} {\it When can information support be useful during in-person group conversations, and how should such support be designed?}
Despite being guided by predefined questions, the discussions were open-ended; participants were encouraged to elaborate on their responses based on their experiences.

\begin{figure}
    \centering
    \includegraphics[width=\linewidth]{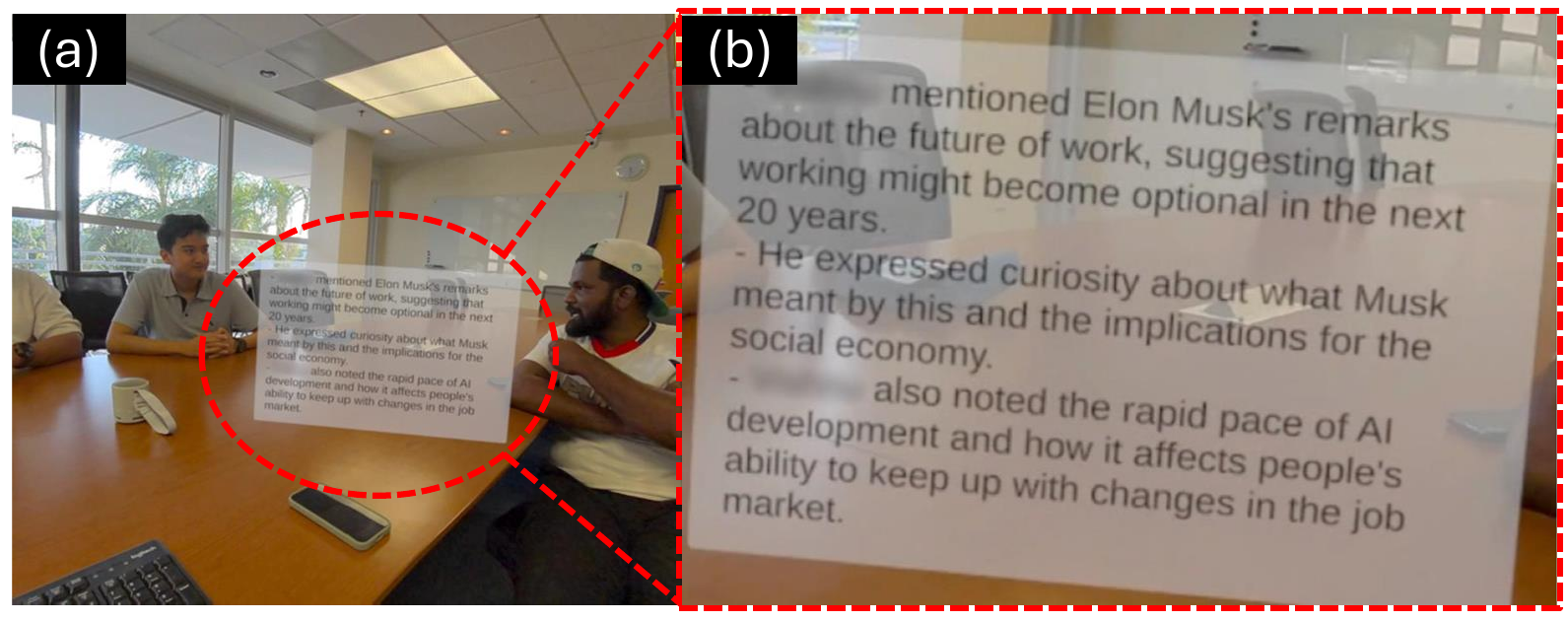}
    \vspace{-0.2in}
    \caption{(a) FPV of provotypes in the formative study. (b) Zoomed-in view of the supporting information.}
    \vspace{-0.2in}
    \label{fig::formative-experience}
\end{figure}

\vspace{2px}\noindent{\bf Phase 3: Participatory Design}.
To understand the design opportunities for proactive information support and the information-seeking behaviors of participants, we developed two \textbf{P}rovo\textbf{T}ypes (PT)~\cite{Boer2012} using the Meta Quest 3S~\cite{metaquest3s} to provoke discussion among stakeholders, including:
\textbf{(PT1)} {\bf a WoZ prototype} that allowed conversation participants to experience proactive information support, with a researcher familiar with the designated topic controlling the supporting information via a dedicated dashboard and presenting it as an MR overlay;
and \textbf{(PT2)} {\bf a MR prototype} that allowed MR users to type questions during in-person group conversations and used GPT-4o-mini~\cite{gpt-4o-mini} to generate supporting information in response to the query input by the MR user.
The support overlay was positioned approximately $70$~cm in front of the user, aligned with the forward direction of their headpose, following Meta's MR design guidelines~\cite{metamrdesign}.

Participants were instructed to experience two provotypes.
Figure~\ref{fig::formative-experience} shows the \textbf{F}irst-\textbf{P}erson \textbf{V}iew~(FPV) of the provotypes.
A focus group discussion was conducted after experiencing each provotype.
Each participant group was instructed to discuss open-ended topics, including {\it where to take a Christmas trip in California} and {\it how AI may reshape the future of work}.
To prevent carry-over effects, topic assignments were counterbalanced across participant groups.

\vspace{2px}\noindent{\bf Data Analysis.}
We used thematic analysis~\cite{Braun2006} and a deductive and inductive coding approach~\cite{Lazar2010} to analyze the transcribed data. Two researchers were involved in the coding process. RQs were used as the basis for deductive coding.
Recorded FPV and \textbf{T}hird-\textbf{P}erson \textbf{V}iew~(TPV) were referenced during data analysis.

\subsection{Findings}\label{sec::formative::results}

Our analysis converged on four themes. 
Participants who experienced PT1 and PT2 are highlighted in \colorbox{mr-tp1}{green} and \colorbox{mr-tp2}{blue}, respectively.
Participants without color highlighting were other group members in the same conversations who did not wear the MR headset and could not see the private supporting information.

\vspace{2px}\noindent{\bf \ul{Affordance of proactive information support.}}
Overall, all participants agreed that the proactive information support helped maintain focus in many conversation scenarios.
A few participants also emphasized the importance and value of real-time support, for example, \textcolor{participant}{\it ``as real-time as possible would be the best!''} (FP2).
First, participants valued {\bf proactive information support when conversations stalled or drifted off topic}. 
\colorbox{mr-tp1}{FP1}, who scored below average in all four aspects of ICSI, commented: \textcolor{participant}{\it ``this is actually a really good thing that while having a conversation, the agent is giving instructions on how to carry the conversation forward.''}
FP2, one of \colorbox{mr-tp1}{FP1}'s conversation partners, mentioned a noticeable change in participation, commenting that \colorbox{mr-tp1}{FP1} was \textcolor{participant}{\it ``generally a bit silent,''} but appeared more engaged, possibly due to the assistance provided by the supporting information.
Participants highlighted the benefit of receiving timely, \insitu~ supporting information, especially when the conversation deviated from the planned topic.
Examples included \textcolor{participant}{\it `` [...] even when I am clueless about what to talk, the instructions were helpful in actually keeping the conversation right to the point because we were actually deviating from the topic and talking about traffic or something''}~(\colorbox{mr-tp1}{FP1}), and \textcolor{participant}{\it ``It’s helpful for meetings in workspace, where you have a goal and want to finish it quickly''}~(FP6).
Second, participants appreciated {\bf receiving suggestions on how the MR user could contribute to the conversation}.
\colorbox{mr-tp1}{FP1} suggested the usefulness of \textcolor{participant}{\it``some important points which we have to consider, but we might forget it at some point.''}
FP4 pointed out: \textcolor{participant}{\it ``[\colorbox{mr-tp1}{FP1}] was better at focusing on the particular topic and asking some interesting questions to make the conversations interesting.''}
\colorbox{mr-tp1}{FP7} highlighted its value for people who feel nervous speaking with strangers:\textcolor{participant}{\it ``I think it definitely helped [...] there's a degree of like nervousness when you're talking to strangers.''}
\colorbox{mr-tp2}{FP5} suggested \textcolor{participant}{\it ``if there is a silence, this can help [...] sometimes you just need to get the talking going.''}
Finally, the focus group highlighted the benefits of {\bf providing additional information during conversations when the topic was less familiar}.
FP4 highlighted that \textcolor{participant}{\it ``it was about Christmas in California. We didn't know much about that. Some [supporting information] can be helpful.''}
In contrast, supporting information presented at less optimal moments could be disruptive, particularly during casual conversations.
FP2 described: \textcolor{participant}{\it ``I felt [\colorbox{mr-tp1}{FP1}] came in between sometimes when we were trying to talk and be casual [...] it's not something relevant to what we were discussing.''}

\vspace{2px}
\noindent
{\bf \ul{Supporting information must preserve, rather than replace, participants' ownership of the conversation.}}
While participants recognized the value of the supporting information overlay, they stressed that it should not give the impression of replacing participants' ownership of the conversation.
For example, \colorbox{mr-tp1}{FP1} noted: \textcolor{participant}{\it ``I feel more comfortable and helpful [...] and rather than being pushed, I can contribute my part of the conversation with that.''}
Participants reported feeling pressured to respond when the supporting information appeared overly directive or when it did not align with their own thought processes.
For example,
\textcolor{participant}{\it ``Even though I know they're just suggestions, I feel pressure to respond and relate to the instructions''} (\colorbox{mr-tp1}{FP7}).
\colorbox{mr-tp1}{FP7} further noted that perceived pressure can increase when a knowledge gap exists between the supporting information and the user, for example: \textcolor{participant}{\it ``it just said ask some activities that can be done during summer, but I don't really know how summer looks like [in California].''}

\vspace{2px}\noindent{\bf \ul{Types of supporting information.}}
During Phase 3, a variety of supporting information was requested by \colorbox{mr-tp2}{FP3}, \colorbox{mr-tp2}{FP5}, and \colorbox{mr-tp2}{FP8}.
We identified four types of supporting information that participants sought: summaries of the prior conversation contexts (\colorbox{mr-tp2}{FP3}, \colorbox{mr-tp2}{FP8}), overlooked aspects of the conversation (\colorbox{mr-tp2}{FP8}), factual clarifications of unfamiliar topics (\colorbox{mr-tp2}{FP5}, \colorbox{mr-tp2}{FP8}), opinions from other conversation participants (\colorbox{mr-tp2}{FP3}).
For example, \colorbox{mr-tp2}{FP3} valued the system's ability to summarize prior conversational context:\textcolor{participant}{\it ``one thing that was most helpful for me is that I was asking the agent to summarize every guest opinion [...] they were giving the summary and during that I did forget his name [refer to FP4].''}
While participants perceived most of the AI-inferred supporting information as useful, our findings revealed several improvements.
\colorbox{mr-tp2}{FP3} suggested shortening the generated prior conversation summary to make it more readable and glanceable on-the-fly: \textcolor{participant}{\it ``the summary is good, but it is a little bit longer [...] I don't know if I'm going to go down or something [\colorbox{mr-tp2}{FP3} refers to reading through it].''}
\colorbox{mr-tp1}{FP7} suggested the usefulness of indicating nonverbal cues of conversation participants due to \textcolor{participant}{\it ``struggling with recognizing sometimes fairly evident social cues.''}

\vspace{2px}\noindent{\bf \ul{Placement of supporting information.}}
Participants reported mixed opinions regarding the visual placement of the MR supporting overlay.
A few participants considered the current design acceptable. For example, \textcolor{participant}{\it ``I like to have it placed wherever I was looking [...] if it is placed on top, maybe we won't be noticing it while having a conversation. So I think it is a perfect placement [...] I think the distance was good''} (\colorbox{mr-tp1}{FP10}).
However, multiple key findings converged across the focus group discussions.
First, a static placement can often cause unnecessary context switching, as users must shift attention between the conversation and the displayed information.
While \colorbox{mr-tp1}{FP1} acknowledged the initial placement, he suggested: \textcolor{participant}{\it ``once it appears in one location, it just stays there. If I look to the side, I can't see it.''}
\colorbox{mr-tp2}{FP3} emphasized the importance of placing the supporting information overlay according to the participant with whom one is conversing: \textcolor{participant}{\it ``the main concern is that the [MR overlay] should be dynamic. It should keep moving [...] For example, if it is fixed over there [\colorbox{mr-tp2}{FP3} pointed to FP2], then I have to take a look at him in order to read what the response is.''}~
\colorbox{mr-tp2}{FP3} suggested that the placement of supporting information should be adapted to the participants who are actively speaking: \textcolor{participant}{\it ``if [the overlay] can move to that person while talking, that could be really awesome [...] let's say if [\colorbox{mr-tp1}{FP1}] is talking, [the supporting information] can move to his direction.''}
Second, some participants noted that the MR overlay can obstruct important nonverbal cues. For example, \textcolor{participant}{\it ``the best way would be to display text on top of the person he is speaking to [...] so you don't break the eye contact''} (\colorbox{mr-tp2}{FP5}).

\subsection{Summary of Design Considerations (DCs)}\label{sec::formative::dcs}

Our findings suggest that while proactive supporting information can be useful, it should augment rather than replace participants' experience of engaging in in-person small-group conversation.
Three DCs are summarized from our formative study.

\vspace{2px}\noindent\fbox{\textbf{DC1}}~{\bf Proactive support should be delivered when conversations stall, drift off topic, or participants lack direction or familiarity with the topic.}
While participants appreciated the helpfulness of proactive support, they emphasized concerns about receiving supporting information when it was not needed.
The design of \sysname~should demonstrate effective ways to control when to deliver proactive supporting information by leveraging heterogeneous verbal and nonverbal conversational context.

\vspace{2px}\noindent\fbox{\textbf{DC2}}~{\bf Supporting information should preserve conversational ownership while integrating verbal and nonverbal contexts to offer summaries, clarifications, overlooked aspects, and relevant others’ perspectives.}
The suggestive information should be context-aware and depend on the conversation participants.
The type of information may vary during different phases of the conversation.
Unlike online conversations (\eg~those outlined in \S~\ref{sec::related::ai_conversation}), support for in-person interactions should account for both verbal and nonverbal behaviors.

\vspace{2px}\noindent\fbox{\textbf{DC3}}~{\bf MR support should be easily consumable during conversation while retaining unobtrusive access to critical nonverbal cues.}
An effective design of \sysname~ should consider the placements of the supporting information overlay.
Given the indispensable role of nonverbal cues in interpersonal communication, the real-time generated MR overlay should not occlude conversation partners while maintaining readability.

\begin{figure}[t]
    \centering
    \includegraphics[width=\linewidth]{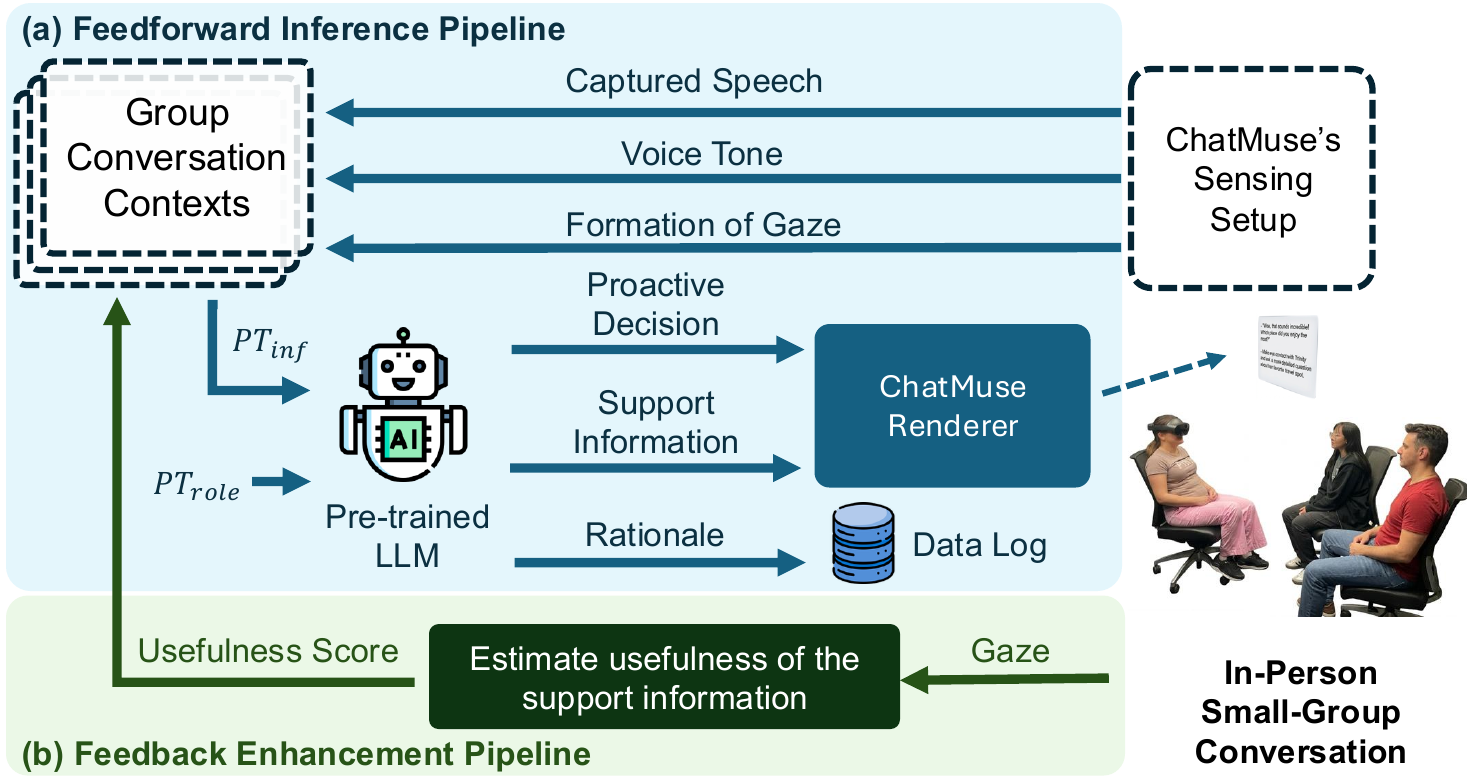}
    \vspace{-0.25in}
    \caption{\sysname's agentic pipeline design.}
    \vspace{-0.2in}
    \Description{place holder}
    \label{fig::architecture}
\end{figure}

\section{\sysname~System Design}\label{sec::system}

Our design of \sysname~is informed by \S~\ref{sec::formative::dcs}.
We consider a collocated small-group conversational setting with three to four participants, in which each participant's headpose can be tracked and speech can be captured (Section \ref{sec::system::setup}).
Figure~\ref{fig::architecture} shows an overview of \sysname~ system, which includes two pipelines:
{\bf (1) a feedforward inference pipeline} that allows the \sysname~agent to generate proactive support in real time by leveraging the verbal and nonverbal behavioral contexts (Figure~\ref{fig::architecture}a, \S~\ref{sec::system::feedforward}), 
and {\bf (2) a feedback enhancement pipeline} that, in turn, uses implicit nonverbal behavior to enhance the \sysname~ agent's ability to assist the MR user in subsequent interactions (Figure~\ref{fig::architecture}b, \S~\ref{sec::system::feedback}).
 
\sysname~ agent leverages a pre-trained LLM to interpret verbal and nonverbal behaviors, decide when to proactively intervene, and generate suggestive text, which is rendered through the MR headset.
Each LLM invocation uses a dedicated prompt, comprising a {\bf role prompt} ($PT_{role}$), an {\bf inference prompt} ($PT_{inf}$).
$PT_{role}$ defines the agent's role and the types of supporting information described in  \S~\ref{sec::formative::results}.
$PT_{inf}$ aggregates key verbal and nonverbal cues from the group conversation context and is continuously updated by the feedforward inference and feedback enhancement pipelines as the conversation progresses.

\subsection{Scenarios and Setups}\label{sec::system::setup}

\begin{figure}[t]
    \centering
    \includegraphics[width=\linewidth]{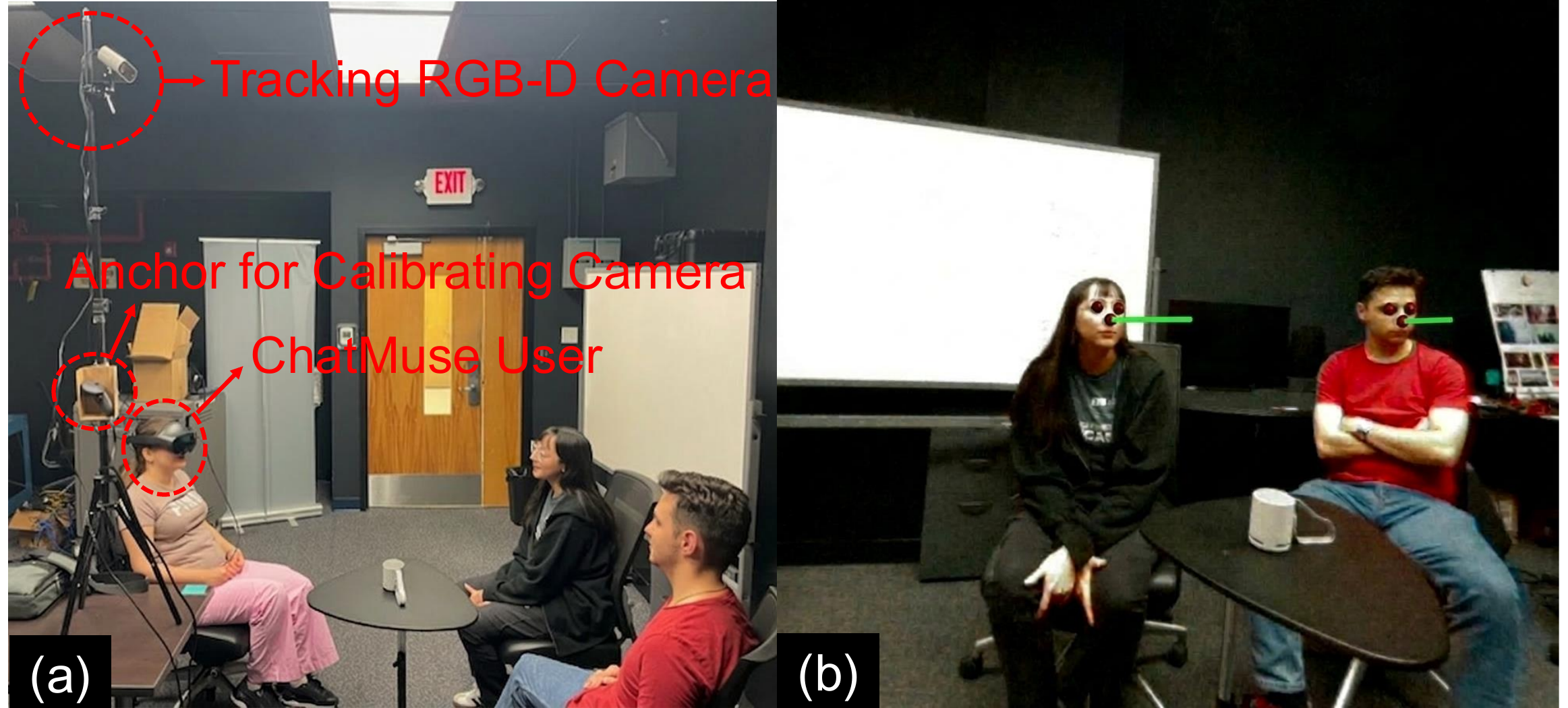}
    \vspace{-0.25in}
    \caption{Setup of \sysname. (a) TPV with three participants; (b) FPV with the forward direction of the tracked headpose visualized}
    \vspace{-0.2in}
    \Description{place holder}
    \label{fig::setup}
\end{figure}

Meta Quest Pro~\cite{metaquestpro} is used due to its ease of prototyping and built-in eye tracking~\cite{metaquestproeyetracking}. 
Although \sysname~ appears to be more effective on current AI glasses, \eg~the recent Ray-Ban Display~\cite{raybandisplay}, today's AI glasses do not support eye tracking.
Our setup includes an Orbbec Femto Bolt camera~\cite{orbbecCam}, an RGB-D camera that supports multi-person pose tracking via Azure Kinect \cite{orbbecCamKinect}, to track all conversation participants other than the MR user.
\sysname~approximates the attention of non-MR users using their forward-facing direction.
While tracking pupil can give an accurate direction of visual attention, estimating the gaze of multiple users remains challenging.
Using headpose to approximate visual attention has been shown to achieve high accuracy in the horizontal direction~\cite{Jha2016}, which aligns with the primary orientation encountered in group conversations.
We also use an external microphone to capture speech, which is then processed through Azure's real-time transcription and speaker diarization services~\cite{AzureSpeechService} to produce timestamped, speaker-labeled transcripts.
Figure~\ref{fig::setup}a shows the setup of \sysname. 
Figure~\ref{fig::setup}b shows the tracked participants from the FPV of the \sysname-supported user.
Appendix~\ref{app::implementation} outlines implementation details.
Appendix~\ref{app::track_performance} presents the benchmarking results for the tracking performance of the RGB-D camera across the pre-designated seating areas.

\subsection{Proactive AI Agent and the Formatting of Support Information}\label{sec::system::support}

\noindent{\bf Design of proactive agent.}
\sysname~uses $PT_{role}$ and $PT_{inf}$ to infer a {\bf proactive decision} and {\bf support information}.
The \emph{proactive decision} is a dedicated flag used to determine whether the \emph{support information} can be a useful addition to current conversation and should be rendered.
While proactive support can benefit MR users, rendering it unnecessarily may introduce unwanted distractions and pressure ({\fbox{DC1}}).
\sysname~provides an opportunity to investigate how effectively today's agents can make proactive decisions by interpreting heterogeneous conversational contexts.

\vspace{2px}
\noindent{\bf Formatting of the support information.}
Structure and formatting of the supporting information are defined by $PT_{role}$.
Guided by \fbox{DC2}, $PT_{role}$ specifies how the agent may provide conversational support, including suggestions for nonverbal behaviors, example phrases the user could say, the appropriate voice tone that the user can follow, and relevant background information to support participation.
Drawing on Yang~\etal's findings~\cite{Yang2025SocialMind} on designing an MR-based dyadic communication support, we structure the support information as bullet points, including example phrases that the MR user can follow.
To ensure the supported user can easily consume the real-time generated conversational support, we empirically limit the output to no more than $50$ words.
Our iteratively refined $PT_{role}$ also incorporates the types of information identified in \S~\ref{sec::formative} while participants were using PT2.

\subsection{Feedforward Inference Pipeline}\label{sec::system::feedforward}

Real-time conversation support should draw on both verbal and nonverbal context to determine what information to surface and when (\fbox{DC1}, \fbox{DC2}).
Interpreting these contextual signals, however, is far from straightforward.
\sysname~addresses this challenge by actively maintaining and updating a \emph{group conversation context} throughout the conversation (Figure~\ref{fig::architecture}a). 
This context is integrated into $PT_{inf}$, which is then analyzed by the AI agent.

\begin{figure}[t]
    \centering
    \includegraphics[width=\linewidth]{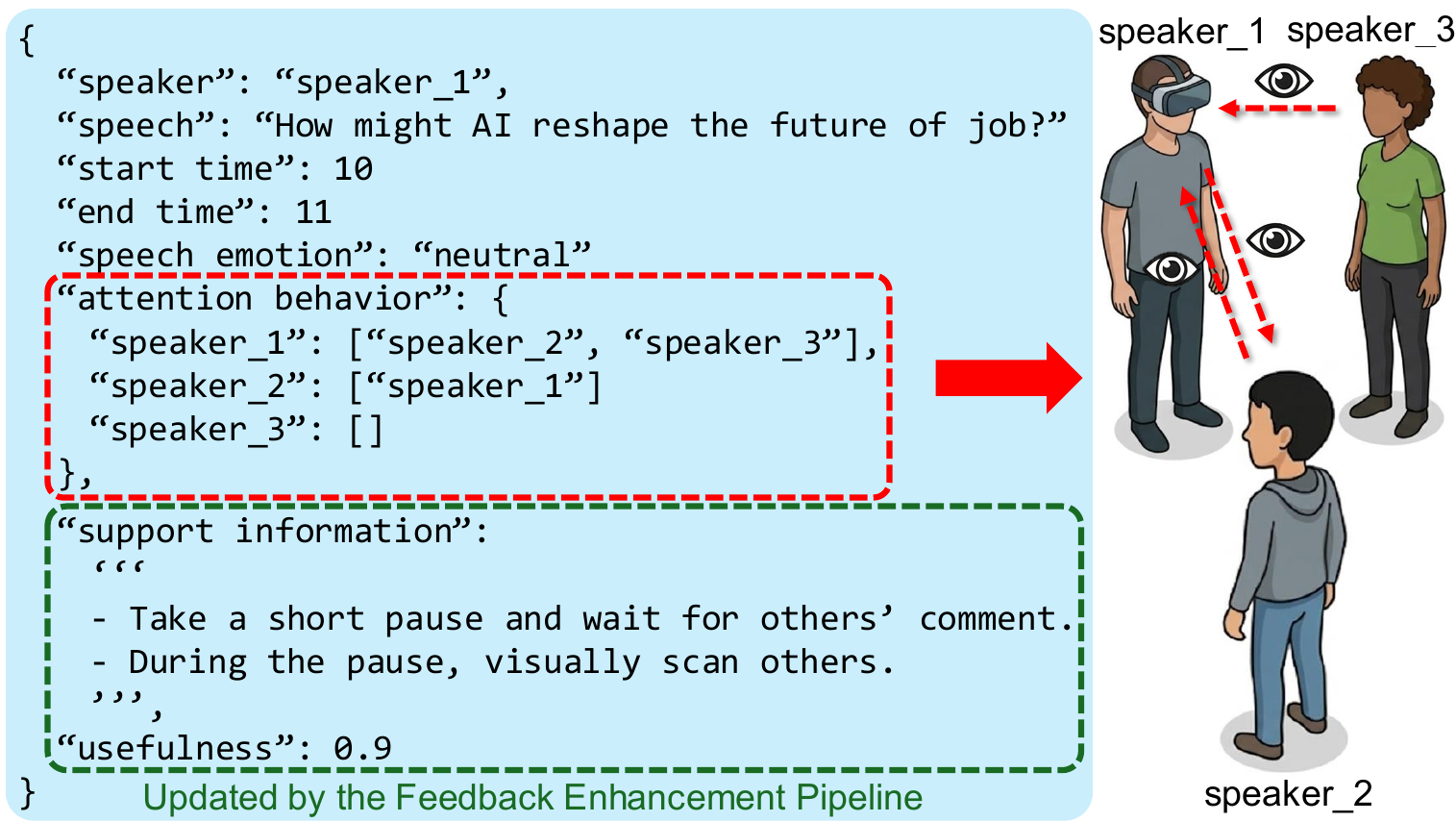}
    \vspace{-0.25in}
    \caption{Demonstrative conversation episode used in the feedforward inference pipeline and subsequently updated through the feedback enhancement pipeline.}
    \vspace{-0.25in}
    \label{fig::episode}
\end{figure}

The {\bf group conversation context} captures both verbal and nonverbal behaviors of all participants throughout the entire conversation.
Unlike group conversations mediated by IM applications (\S~\ref{sec::related}), both verbal and nonverbal cues are essential in \emph{in-person} group interactions (\fbox{DC2}).
Accordingly, \sysname~incorporates \emph{speaker}, \emph{verbal speech}, \emph{speech emotions}, \emph{start and end timestamp}, and \emph{formation of attention} for each {\bf conversation episode}, which is defined as the tracked contextual data while one participant is speaking (Figure~\ref{fig::episode}).
While we recognize the significance of other nonverbal cues, such as hand gestures~\cite{Maricchiolo2011, Krauss1995, Okada2013} and f-formations~\cite{Marshall2011, Yoo2026}, the current implementation of \sysname~focuses solely on incorporating attentional behaviors.
Nevertheless, this represents a first step toward exploring agent-supported, MR-based in-person small-group conversation experiences, with future work aimed at integrating additional nonverbal behaviors.
Although most of the contextual data in each conversation episode is straightforward, we describe how speech emotions and the formation of attention are captured and integrated into the group conversation context, and thereby $PT_{inf}$.

\vspace{2px}\noindent$\bullet$~{\bf Inference of speech emotion.}
\sysname~captures the raw audio data for each conversation episode, which is then used to infer the emotions expressed in the speech.
We focus exclusively on the six basic emotions proposed by Ekman~\etal, alongside the neutral emotion~\cite{Paul1992, Ekman1971}.
We use the pretrained speech emotion classification model~\cite{Sakthi2025}, which achieves an overall accuracy of $83.79\%$ on benchmark datasets.

\vspace{2px}\noindent$\bullet$~{\bf Formation of attentions.}
Leveraging the tracked pose data of all conversational participants, we represent the formation of attention as a directed graph, in which each conversational participant is modeled as a node, and the directed edges between nodes encode the looking directions among participants.
Each node may have zero inward edges, indicating that the corresponding participant is not looked at by any other participant, or multiple inward edges, indicating that the participant is looked at by more than one participant during the associated utterance.
This graph is then represented as a \texttt{JSON} object, where each key corresponds to a node mapped to a list of nodes that it is looked at by.
Figure~\ref{fig::episode} illustrates an example of the formation of attention.
For simplicity, we assume that a conversational participant (\eg~ \emph{speaker\_1}) is being looked at by another participant (\eg~ \emph{speaker\_2}) only when the horizontal angle between \emph{speaker\_2}'s gaze direction and the direction vector from \emph{speaker\_2} to \emph{speaker\_1} is less than $30^\circ$.

\subsection{Feedback Enhancement Pipeline}\label{sec::system::feedback}
While the feedforward inference pipeline relies solely on prior conversational context, the feedback enhancement pipeline additionally leverages MR users' nonverbal behaviors following the rendering of supporting information to improve future support generation (Figure~\ref{fig::architecture}b).

To achieve this, \sysname~ estimates a {\bf usefulness score} ($u$) after rendering each supporting information, reflecting how effectively it assisted the MR user's conversation experience.
The estimated $u$ and the associated generated supporting information are then updated in the corresponding conversation episode (Figure~\ref{fig::episode}).

While visual attention does not directly indicate usefulness, its allocation in task-driven contexts is heavily influenced by task relevance~\cite{duchowski2017}. Gaze duration is commonly used as a proxy for visual attention, as longer viewing time reflects sustained attentional allocation to a stimulus~\cite{Just1980}. We compute $u = \frac{t}{T}$, where $t$ is the user's gaze duration on the MR support overlay and $T$ is the expected reading time of the generated text.
For simplicity, $T = \frac{60N}{wpm}$, where $N$ is the number of generated words, and $wpm$ is the typical adult reading speed.
We set $wpm = 238~words/min$ based on empirical data on the silent reading speed of non-fiction texts for an adult~\cite{Brysbaert2019}.

\subsection{Rendering Virtual Support Information}\label{sec::system::placement}

\begin{figure}
    \centering
    \includegraphics[width=\linewidth]{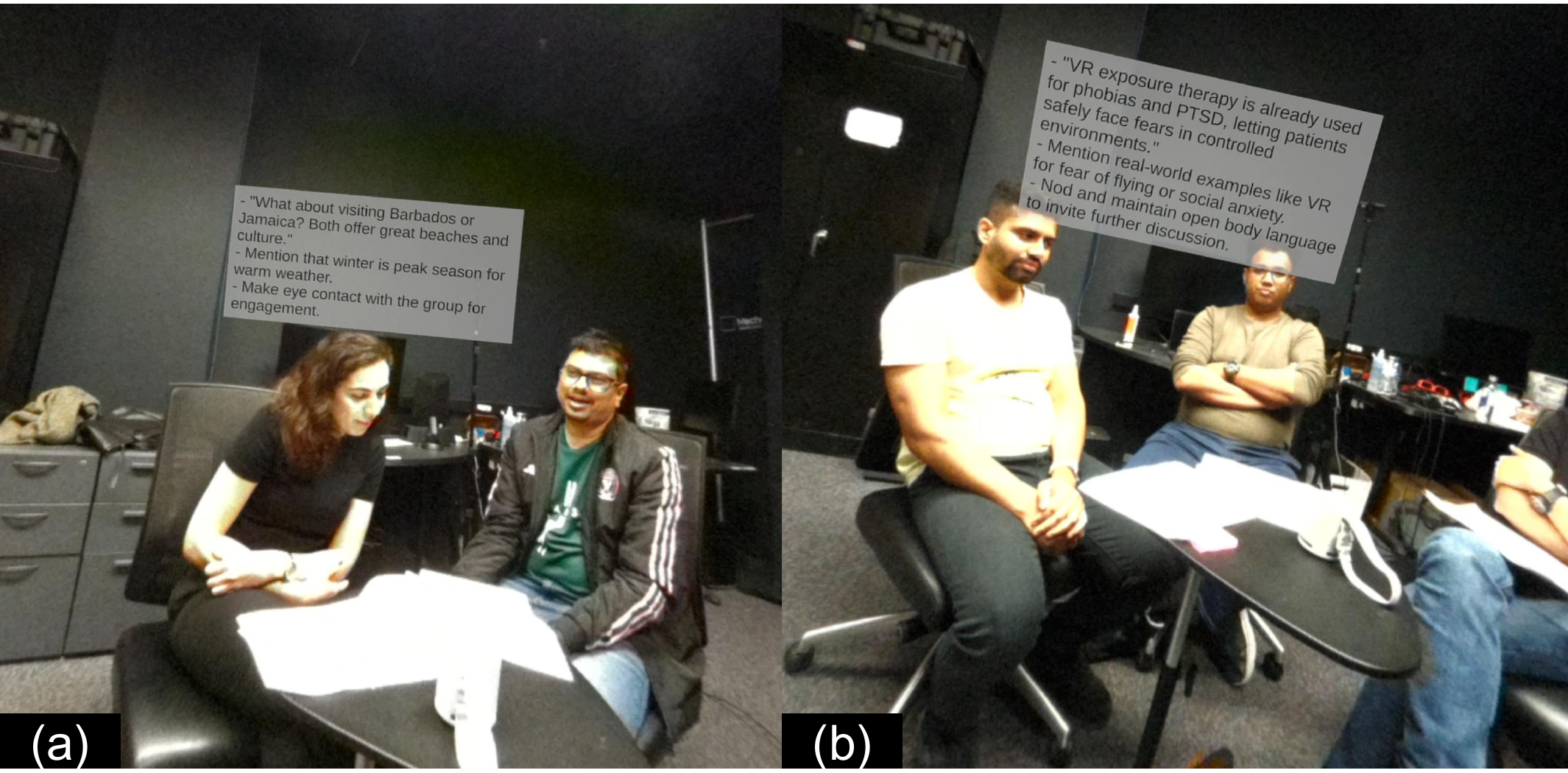}
    \vspace{-0.25in}
    \caption{Placements of virtually rendered support within FPV when (a) the supported user is looking at one conversation partner and (b) when the supported user is not.}
    \vspace{-0.2in}
    \label{fig::placement}
\end{figure}

Informed by \fbox{DC3}, the rendered support information should remain readable without occluding other conversation partners.
To address this DC, we consider two scenarios.
When the inferred support content explicitly mentions a specific conversation participant whom the supported user should refer to, the rendered support content is placed above that participant's head.
Otherwise, \sysname~determines the placement based on the MR user's gaze behavior.
More specifically, when the MR user looks at another conversation participant, \sysname~positions the virtual overlay above that participant's head (Figure~\ref{fig::placement}a).
In contrast, when a participant does not look at their conversational partners, we simply place the overlay in front of the MR user along their gaze direction (Figure~\ref{fig::placement}b).

\section{Evaluation}\label{sec::evaluation}

We conducted within-subject studies, alongside focus groups, to understand the feasibility, utility, and usability of \sysname.
More specifically, we aim to address three RQs:
{\it how does real-time conversational support information influence participants' engagement and contributions during in-person small-group conversations?}~\textbf{(RQ1)}, {\it how does the feedback enhancement pipeline affect \sysname's ability to support conversation participants?}~\textbf{(RQ2)}, and {\it to what extent does the placement of the conversation supporting MR overlay provide support to conversation participants without causing distractions or occlusions?}~\textbf{(RQ3)}

\subsection{Participants and Procedures}
\noindent {\bf Participants.} 
We recruited $18$~participants (age, $M = 26.1$, $SD = 4.0$) and organized them into six groups: five groups of three and one group of four (Table~\ref{tab::participant}, Appendix~\ref{app::participants}).
P18 withdrew from the study due to an unexpected event. As a result, one of the researchers served as the unsupported conversation participant, but did not participate in or influence the post-study focus group discussion.
P\# and G\# are used to indicate \textbf{P}articipants and participant \textbf{G}roup. Participants wearing MR headset are highlighted by \colorbox{mr}{orange}.
All participants reported prior experience with AI tools, though their frequency of participation in group conversations varied.

\vspace{2px}\noindent{\bf Tasks and conditions.}
Participants were first introduced to the study.
In each group, one participant was assigned as the MR user, while the remaining participants served as conversation partners.
Each group was then instructed to complete the training conversation session, which allowed participants to become familiar with the study setting and the MR interface.
The MR user was instructed to calibrate the eye tracking and all participants were asked to calibrate the tracking system at the start of each session.
The participant group was then asked to complete three supporting conditions, with the order counterbalanced using a Latin square design~\cite{Fisher1950LatinSquare}.

\vspace{2px}
\begin{itemize}[topsep=0pt, itemsep=0pt, leftmargin=*]

  \item {\bf Baseline ($C_{baseline}$)}. The MR user wore the headset during the conversation with only passthrough enabled, without \sysname's support. This baseline condition allowed us to isolate the effect of wearing the MR headset.

  \item {\bf Support without Feedback Enhancement ($C_{NoFeedback}$)}. While the \sysname~ system was deployed in this supporting condition, the feedback enhancement pipeline was disabled.

  \item {\bf \sysname~system with feedback enhancement ($C_{\sysname}$)}. \sysname~ system was used in this condition. In this setup, both the feedforward inference pipeline and the feedback enhancement pipeline are employed to control and generate real-time conversational support.

\end{itemize}

During each session, participant groups were directed to engage in discussions on three topics: \textbf{(T1)} {\it how will AI affect future careers and the way people work}; \textbf{(T2)} {\it if we were to travel to the Caribbean next winter break, where should we go}; \textbf{(T3)}~{\it what are the promising VR applications in the healthcare domain}. 
These topics were iteratively selected by our team because they are open-ended and require a certain level of domain knowledge.
Such topics could potentially introduce friction in group conversations.
Each conversation session lasted around $15$~min.

After each session, participants were instructed to complete the \textbf{NASA} \textbf{T}ask \textbf{L}oad Inde\textbf{x} (NASA TLX)~\cite{Hart1988NASATLX} to evaluate their perceived workload.
Participants wearing the MR headset were asked to complete an additional \textbf{S}ystem \textbf{U}sability \textbf{S}cale (SUS)~\cite{Brooke1996sus} questionnaire to assess their perceived usability. 
At the end of the sessions, MR users were asked to rate five statements (Q1--Q5) on a five-point Likert scale, while non-MR users rated only Q1--Q3.
The statements for Q4 and Q5 did not reference a specific condition; instead, participants were prompted to reflect on either their first, second, or third session. We explicitly include the condition to improve readability.

\begin{itemize}[topsep=0pt, itemsep=0pt, leftmargin=*]
    \item \textbf{Q1:}~{\it ``Having conversational support helps me better engage in and contribute to the group conversation.''}
    \item \textbf{Q2:}~{\it ``Having conversational support makes me feel more overwhelmed.''}
    \item \textbf{Q3:}~{\it ``The support information helps me recall prior points from our group conversation.''}
    \item \textbf{Q4:}~{\it ``In terms of the usefulness of the suggestive content, $C_{\sysname}$ is more helpful than $C_{NoFeedback}$.''}
    \item \textbf{Q5:}~{\it ``In terms of the proactivity, $C_{\sysname}$ is more helpful than $C_{NoFeedback}$.''}
    
\end{itemize}

Finally, we conducted a focus group to gain a deeper understanding of participants' experiences.

\subsection{Data Analysis}
We analyzed the responses from post-study questionnaires and the focus-group transcripts.
The Friedman test was used to evaluate the quantitative questionnaire responses ($\alpha = .05$)~\cite{Friedman1937}.
Thematic analysis~\cite{Braun2012} and deductive and inductive coding~\cite{Elo2008} were used to analyze qualitative data.
RQs were used to establish the initial themes.
Inductive coding~\cite{Elo2008} was used to analyze the support content and the rationales inferred by the AI agent. 
While our analysis primarily focused on qualitative data, we also computed three objective measures: 

\begin{itemize}[topsep=0pt, itemsep=0pt, leftmargin=*]
    
    \item {\bf proactive rate} of the AI agent. The proactive rate was computed as the ratio of decisions in which the agent determined it should provide support to the total number of agent invocations.

    \item  {\bf participation inequality}, capturing the dispersion in contributions. We first measured the \% of total speaking time contributed by each participant. Participation inequality was then estimated as the standard deviation of these proportions.

    \item {\bf support adoption score}, measuring the extent to which MR~users incorporated real-time support content. Support adoption score for a specific support information was approximated by the maximum cosine similarity between the textual embeddings of the example speech and the utterances produced by the \sysname~user prior to the subsequent support content being rendered. Pre-trained sentence transformer \texttt{all-MiniLM-L6-v2} was used to encode textual semantics \cite{sentence-transformer, Reimers2019}. We employed Tau-U analysis ($\alpha = .05$)~\cite{Parker2011TauU} to evaluate the non-overlap in support adoption scores between $C_{NoFeedback}$ and $C_{\sysname}$.
    
\end{itemize}

\subsection{Results}

Our results were organized across the three RQs.
Our findings were further grounded in the analysis of $209$ and $271$ textual rationales generated by $C_{NoFeedback}$ and $C_{\sysname}$.
Table~\ref{tab::participation} reports the participation inequality and support adoption scores for each group.
To improve readability, we used \textcolor{participant}{gray} to highlight participants' testimonies and \textcolor{agent}{blue} to highlight agent-inferred support content and accompanying textual rationales. 
Appendix~\ref{app::survey} presents the SUS and NASA TLX scores.

\vspace{2px}\noindent{\bf \ul{How does ChatMuse influence participants' engagement and contributions?~(RQ1)}}

\begin{figure}[t]
    \centering
    \includegraphics[width=\linewidth]{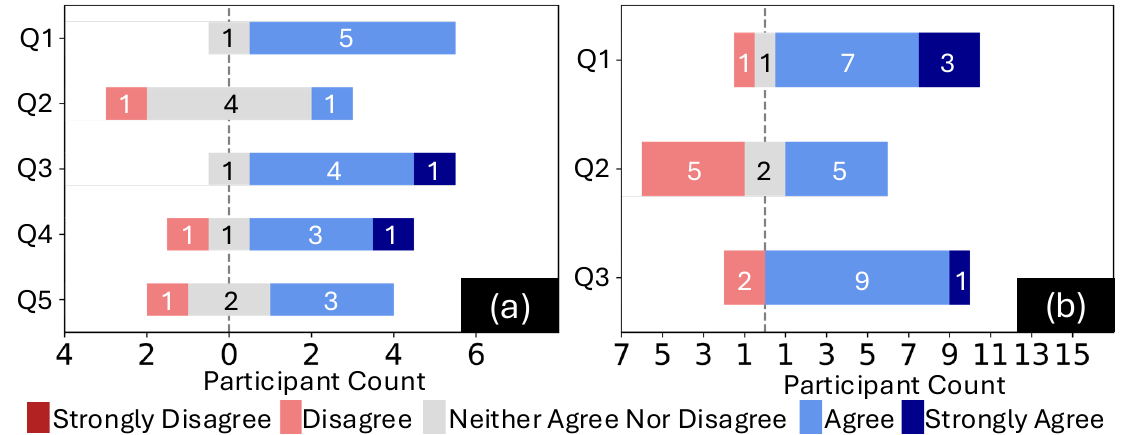}
    \vspace{-0.22in}
    \caption{Post-study survey responses from (a) MR users and (b) non-MR users.}
    \vspace{-0.22in}
    \label{fig::seq-questionaire}
\end{figure}

\noindent Most participants valued the usefulness of proactive conversation support in assisting their participation in group discussions.
Most participants viewed \sysname~ as an \textcolor{participant}{\it ``assistive tool or assistive agent''}~(\colorbox{mr}{P17}) for small-group conversation.
Although no significant differences were observed in the SUS and NASA TLX responses, the benefits were reflected in the responses of Q1 (\mbox{Figure~\ref{fig::seq-questionaire}a--b}).

\vspace{2px}
\noindent {\bf \emph{What key benefits does \sysname's support provide?}}
Overall, we identified seven themes representing the key benefits: {\bf reduced conversational friction} (P5, \colorbox{mr}{P7}, \colorbox{mr}{P11}), {\bf support in providing summarized context} (\colorbox{mr}{P4}), {\bf access of essential background information} (\colorbox{mr}{P7}), {\bf support in starting a new topic}~(\colorbox{mr}{P7}, \colorbox{mr}{P17}), {\bf emphasizing the context of the conversation}~(\colorbox{mr}{P17}), {\bf assistance in extending current discussion topics} (\colorbox{mr}{P14}), and {\bf serving as the source of inspirations} (\colorbox{mr}{P14}).
For example, \colorbox{mr}{P1} appreciated the possible speech generated by \sysname~ that he could use in conversation: \textcolor{participant}{\it ``the example line is helpful [...] it was giving me instructions regarding what to engage and that kind of stuff too [...] It was helpful!''}
The focus group of G3 highlighted benefits such as reducing conversational friction and using the system as an icebreaker.
\colorbox{mr}{P7} pointed out: \textcolor{participant}{\it ``the supports not only help me to initiate the conversation and a new topic, but also probe the critical background.''}~
After P9 remarked: \textcolor{participant}{\it ``the conversation flowed well, and I gained many ideas about healthcare,''} \colorbox{mr}{P7} immediately noted: \textcolor{participant}{\it ``but I was the one who started it, right? Otherwise we would have all been stuck.''}
Some participants also highlighted the usefulness of providing additional background information.
Examples include:
\textcolor{participant}{\it `` I think it's pretty good in the last session because I don't know much about healthcare, and the suggestions provided were useful''} (\colorbox{mr}{P1}); 
and \textcolor{participant}{\it ``I didn't have a lot of knowledge in [T2]. So it was helping kind of provide topics to talk about''} (\colorbox{mr}{P4}).
\colorbox{mr}{P14} emphasized the value of using the provided background to extend the argument: \textcolor{participant}{\it ``[the support] can be useful to help me extend my perspective. For example, for T2, while I know a few things about the Caribbean, like the cruise, it provides me with some examples to help me start the conversation, and try to get everyone involved.''}
Further, participants also highlighted the benefits when AI-generated support serves as the source of inspiration: \textcolor{participant}{\it ``But when I try to get deeper and try to talk more about some specific area, I really appreciate the clues which get to inspire me to give me more cues and options''}~(\colorbox{mr}{P14}).
Finally, some participants appreciated the potential benefit of having access to a conversation summary: \textcolor{participant}{\it ``I really appreciate the summary at the end because sometimes I'll just be talking and I don't remember what we just talked''}~(\colorbox{mr}{P17}).

\vspace{2px}\noindent{\bf \emph{How is the proactive support perceived?}}
\noindent Participants valued the proactive support offered by \sysname, but differed in how often they wanted it to be provided.
%
Example include: \textcolor{participant}{\it ``[the support] updates by itself, it's far good [...] the conversation flew very nicely, mostly because of the automatically pop-up support actually''}~(\colorbox{mr}{P7}); and \textcolor{participant}{\it ``it was helpful that the suggestions could automatically update based on our conversation''}~(\colorbox{mr}{P11}).
A few participants, however, noted that the timing of rendering the support content could be further improved.
\colorbox{mr}{P1} believed that \textcolor{participant}{\it ``sometimes, the support message keeps changing too frequently,''} although \colorbox{mr}{P14} thought \textcolor{participant}{\it ``[frequently update of the support] doesn't matter because I have some word in my mind. With the suggestion, I am just trying to find a good time to cut it in.''}
On the contrary, \colorbox{mr}{P7} mentioned that \textcolor{participant}{\it ``at some points, I was very worried why [the support] is not coming.''}
Our analysis of the AI-generated support content reveals proactive rates of $87.98\%$ ($SD = 8.57\%$) for $C_{NoFeedback}$ and $89.91\%$ ($SD = 6.83\%$) for $C_{\sysname}$.
We observed that the AI agent often chose \emph{not} to provide support to avoid disrupting other participants' thought processes (\eg~\textcolor{agent}{\it ``P15 has taken the floor and is introducing a new point about Upwork. It is appropriate to listen and wait before contributing further, so no support is needed at this moment''})~(G5, $C_{\sysname}$), 
to prevent intervening when the MR user was not actively needed in the conversation (\eg~\textcolor{agent}{\it ``P8 is currently speaking and responding to \colorbox{mr}{P7}'s question, so it is appropriate for \colorbox{mr}{P7} to listen and wait for P8 to finish before contributing further''})~(G2, $C_{\sysname}$), when the MR user is over active (\eg~\textcolor{agent}{\it ``\colorbox{mr}{P14} has just provided a thorough and wide-ranging answer, and another participant has complimented the comprehensiveness. No immediate support is needed as \colorbox{mr}{P14} is actively engaged and has completed its contribution''})~(G4, $C_{NoFeedback}$), and when a participant was attempting to summarize the discussion (\eg~\textcolor{agent}{\it ``P19 is currently summarizing the conversation, and \colorbox{mr}{P17} has already provided a comprehensive summary. No additional support is needed at this moment for \colorbox{mr}{P17} to engage or contribute further''})~(G6, $C_{NoFeedback}$).

\vspace{2px}\noindent{\bf \emph{How is \sysname's support used?}}
\noindent Our analysis revealed that participants expressed diverse perspectives on the overall perceived usefulness of the support, the content of the textual support, and the ways in which the support was used.
Participants reported mixed experiences with the content supported by AI.
Most MR users found the support information helpful.
For example, \colorbox{mr}{P14} explained: \textcolor{participant}{\it ``the current support is very good because I have full options. I can just choose whether to read it or just use some keywords from the instructions. If I was wrong, I can just use the full instruction again and try to change the subject, try to correct it, and then just express again.''}
However, a few participants identified areas for improvement.
\colorbox{mr}{P4} mentioned experiencing some cognitive load when reading the support content and converting it into her own words: \textcolor{participant}{\it ``when I see the support, I found that I need to kind of translate it into my own way [...] I think maybe just making a bullet point of related topics - that would be much more helpful!''}
\colorbox{mr}{P1} suggested improving the system by making the supporting text more concise: \textcolor{participant}{\it ``I can read the text better if it's shorter.''}
However, \colorbox{mr}{P17} believed that \textcolor{participant}{\it ``since it's more of something that I can just receive and choose to either read off of or choose to dismiss, it's more than welcome to appear.''}~
\colorbox{mr}{P14} appreciated the usefulness of suggestions related to nonverbal cues, although he noted that he did not always follow these suggestions in practice: \textcolor{participant}{\it ``my style to express myself is to have some body language, not as facial expressions [...] So basically [the support] just reminds me to get them [refer to other participants] into this discussion.''}~

When asked about the performance of supported participants during the experimental conversation, most non-MR users reported that they \textcolor{participant}{\it ``could barely tell the difference''}~(P19).
However, these participants, in some contexts, described the MR users as the \textcolor{participant}{\it ``presenter''}~(P5), \textcolor{participant}{\it ``leading role''}~(P6), \textcolor{participant}{\it ``narrator''}~(P8), as well as being \textcolor{participant}{\it ``artificial while transitioning topics''}~(P13), \textcolor{participant}{\it ``giving a speech''}~(P16), and \textcolor{participant}{\it ``less genuine''}~(P16).
A few participants were concerned that the support content may significantly alter the flow of the group conversation.
In the focus group of G4, non-MR participants reflected their experience during the discussion of T3: \textcolor{participant}{\it ``you were talking about places like Cuba [...] And he [\colorbox{mr}{P11}] just says that the Bahamas is really sunny and nice''} (P13); meanwhile, P12 followed up \textcolor{participant}{\it ``it was trying to like veer back from going off topic, but I think the way it did that was like very sudden.''}

\vspace{2px}\noindent{\bf \ul{How does feedback enhancement pipeline affect ChatMuse's ability to support conversation participants? (RQ2)}}

\noindent 
Our findings demonstrate how the feedback enhancement pipeline influences the dynamics of group conversation  and shapes participants' perceptions of the supported experience.

Table~\ref{tab::participation} shows a reduction in participation inequality across all groups for $C_{\sysname}$ compared to $C_{NoFeedback}$.
Significant Tau-U effects~($p < .05$)~for G3--G6 showed an increase in support adoption scores when feedback was provided.

Q4 and Q5 in Figure~\ref{fig::seq-questionaire} summarize participants' post-study survey responses regarding the usefulness of suggestive content and helpfulness of the proactive support.
\colorbox{mr}{P7} and \colorbox{mr}{P11} responded positively\footnote{We consider a response as \emph{positive} when the participant selects agree or strongly agree in Figure \ref{fig::seq-questionaire}.}, preferring $C_{\sysname}$ over $C_{NoFeedback}$ in terms of both the usefulness of the suggested content and the proactivity in presenting it.
Although \colorbox{mr}{P1} neither agreed nor disagreed with Q4 and Q5, he remarked: \textcolor{participant}{\it ``[$C_{\sysname}$] seems to provide more information.''}
While \emph{disagreed} with both Q4 and Q5, \colorbox{mr}{P4} explained her experience: \textcolor{participant}{\it ``in [$C_{NoFeedback}$], I think what it was at was very centered on one point that was made, whereas [for $C_{\sysname}$], it offered different information like variety.''}
On the other hand, the majority of participants did not perceive a difference between $C_{\sysname}$ and $C_{NoFeedback}$ conditions.

Analysis of the rationales inferred by the AI agent suggested that the feedback enhancement pipeline improved the conversational grounding of the generated support.
In $C_{\sysname}$, rationales were frequently tied to what a participant had just said or done in the conversation. 
For example, \sysname~ reasoned that \textcolor{agent}{\it `` \colorbox{mr}{P4} expressed uncertainty about Aruba's location''}~(G2), \textcolor{agent}{\it ``P8 just contributed a thoughtful example''}~(G3), or \textcolor{agent}{\it ``P12 is questioning the use of VR for phobia treatment''}~(G4), and then generated support tailored to those specific moments.
\sysname~ also reflected on whether a prior support strategy had been useful. For example, one rationale stated, \textcolor{agent}{\it ``Prior support with facts was useful, so a similar approach is warranted''}~(G4), showing that \sysname~ used earlier interaction feedback to justify a follow-up intervention with similar approach.
In contrast, rationales in $C_{NoFeedback}$ condition tended to be framed at the broader topic level, such as \textcolor{agent}{\it ``the conversation is focused on safe and stable Caribbean destinations''}~(G1), or \textcolor{agent}{\it ``the conversation is focused on AI's impact on education and training''}~(G2), which simply helped keep the conversation going or broaden the current topic. 

\begin{table}[t]
    \centering
    \caption{Participation inequality and support adoption score for each participant group. Significant Tau-U indices~($p < .05$) are highlighted in \colorbox{green!20}{green}; non-significant in \colorbox{red!20}{red}.}
    \vspace{-0.12in}
    \includegraphics[width=\linewidth]{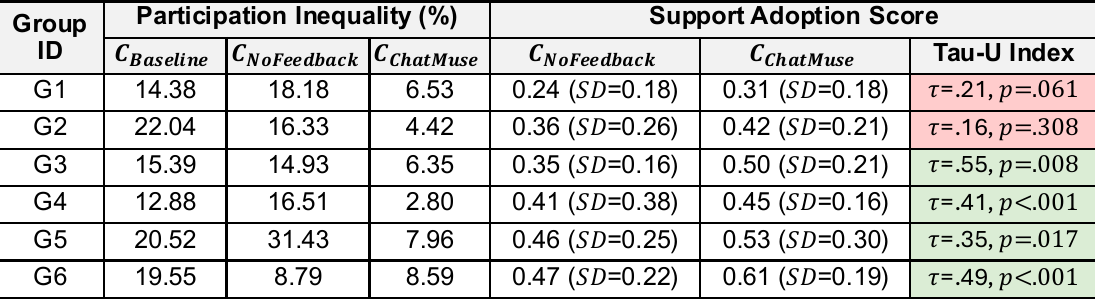}
    \vspace{-0.20in}
    \label{tab::participation}
\end{table}

\vspace{2px}\noindent\textbf{\ul{Placement of the rendered conversation support (RQ3).}}
\noindent Most participants were satisfied with how the conversation support was displayed (\eg~\textcolor{participant}{\it ``if I'm looking at him, the box is right there!''}~(\colorbox{mr}{P1})).
For non-MR users, most participants reported that they were unable to perceive the diverse and unexpected behavioral gestures of participants wearing the MR headset, for example, \textcolor{gray}{\it ``I didn’t feel that he was making any odd head movements''}~(P15).~

However, a few participants wearing MR headset pointed out the limitations of the rendered textual support when it cannot keep pace with rapid changes of attention.
\colorbox{mr}{P7} and \colorbox{mr}{P14} noted the drawback when the supporting text lagged behind the dynamic gestural behaviors.
For example, \textcolor{participant}{\it ``when I start looking at [P9] while speaking, [the support] would be here [...] I really want to look at him [...] So if it actually moves on him with a transparent view, it'd be very good!''} (\colorbox{mr}{P7}).
Although G2 did not encounter this situation, \colorbox{mr}{P7} anticipated a more complex scenario when \textcolor{participant}{\it ``two [or more] people are talking.''}
\section{Discussion}\label{sec::discussion}
\subsection{Practical Implications}\label{sec::discussion::implication}

Our research around \sysname~shows a first design exploration of a new MR-based conversational support. We discuss how our findings inform the design of future MR conversational support agents by incorporating richer contextual information and innovating strategies to preserve conversational ownership while providing on-the-fly \insitu~support.
We then discuss how \sysname~can be potentially integrated into a broader range of applications and support diverse user populations.

\vspace{2px}\noindent{\bf Supporting conversations through the careful integration of richer contexts.}
We demonstrate the effectiveness of \sysname~in supporting group conversations by reducing a variety of conversational frictions, providing summaries, and aiding in the recall of prior context, as well as offering additional background information and topic suggestions.
However, participants also emphasized the importance of incorporating richer conversational context and designing more personalized support.
For example, \sysname~could construct a user profile that represents the supported individual's personas. 
Incorporating these personas may enable \sysname~to adapt its output based on the user's personality and prior knowledge; for example, a lower level of proactive support may be provided to users with greater knowledge of the discussion topic, while more support for gestural behaviors could be designed for users with less developed interpersonal communication skills.

\vspace{2px}\noindent{\bf Conversation ownership.}
Support from \sysname~ should act as a conversational assistant rather than a script for the user to simply read.
The role of \sysname~is to help users stay engaged, notice possible entry points, and sustain the interpersonal relationship with partners, while leaving the final decision of what to say and how to say it to the user.
While our research around \sysname~ explored a new MR-based supporting technique for in-person small-group conversation, our findings revealed distinct perspectives from \sysname~users regarding the perception of rendered content, the timing of textual support delivery, and the use of \sysname-generated support.
While most participants believed that \sysname~ could support their conversational experience, a few (\eg~\colorbox{mr}{P4}) felt overwhelmed at certain moments.
Our findings suggest that a consistent level of agency may not be suitable for all \sysname~users; the agent's level of agency may need to be dynamically adapted based on the evolving context and different stages of group conversation.
With \sysname, a natural next step is to examine how AI-mediated group conversations shape users' sense of ownership, drawing on the construct of locus of control and preference of agency that captures individuals' perceived agency over events~\cite{Rotter1966, Shneiderman2000, Caliendo2024}.
Although the locus of control of AI agents has been studied in other AI-mediated tasks such as writing~\cite{Jakesch2023, Fu2025}, it remains underexplored in our group conversation context.
Future designs of \sysname~ may consider enabling the AI agent to assess the cognitive load of all conversation participants, potentially measured through gaze patterns~\cite{Stiber2024} and pupil dilation~\cite{Gavas2017}.

\vspace{2px}\noindent{\bf Integration with a broader range of applications and users.}
We demonstrate the effectiveness of \sysname~ in supporting in-person small-group conversations.
While the study examines scenarios involving three to four participants across three designated topics, its practical implications extend to a wider range of applications and user contexts.
Our study can be generalized to broader 
real-world group conversation scenarios.
For example, \sysname~ can support individuals with verbal communication challenges or experiences of social withdrawal in more effectively engaging in small-group conversations~\cite{Rubin2001}.
In educational settings, \sysname~ may be applied in active learning classrooms, where small-group discussions are considered a critical component~\cite{Young2013, Keeler1995}.
\sysname~may be valuable for general and domain-specific knowledge workers.
For example, \textcolor{participant}{\it ``I would love to use it for supporting research discussions with my colleagues''}~(P3), and \textcolor{participant}{\it ``I'm doing senior design, ideas that were just spitballing and just coming up with like on the fly [...] I think that [\sysname] would definitely be helpful to be used in such a group discussion!''} (\colorbox{mr}{P17}).

\subsection{Limitation and Future Work}\label{sec::discussion::limitation}

We summarize our limitations in six aspects.

\vspace{2px}\noindent{\bf System constraints.}
We first acknowledge that our study does not fully capture the authentic experience of conversation partners, as the bulky headset form factor and relatively low-quality video passthrough of the Meta Quest Pro~\cite{metaquestpro} can obscure the supported user's eyes and upper facial expressions.
Although our primary focus is the experience of the \sysname~user, the experiences of conversation partners can, in turn, affect the supported user's experience.
We chose the Meta Quest Pro~\cite{metaquestpro} for its built-in eye-tracking capabilities and ease of prototyping.
We believe that $C_{Baseline}$ can isolate confounds introduced by the headset. Future work may investigate the usability and utility of \sysname~when being integrated on lightweight AR glasses, which may provide conversation partners with greater visual access of the eye and facial expressions of the supported user.
Second, the existing Orbbec camera~\cite{orbbecCam} limits nonverbal behavior capture fidelity; more precise motion capture setup like OptiTrack~\cite{OptiTrack} could improve reliability.

\vspace{2px}\noindent{\bf Interaction design and rendering of support.}
Our current interaction design is not optimal.
The current design of textual support overlays may create visual distractions of the MR user.
Future work may extend \sysname~by incorporating the design of prior MR notifications~\cite{WearableDisplay, Janaka2022, Cai2023} research, while optimizing the placement and content of the rendered MR textual overlays.
While \sysname~currently uses a rule-based approach to determine the placement of textual support (Section~\ref{sec::system::placement}), future work may investigate how \sysname~may integrate the design of context-aware~\cite{Fischer2012} placement algorithms based on constrained optimization~\cite{Chen2023, Liang2021, Lang2019, Davari2024, Cheng2021, Nguyen2024PaperToPlacePatent} so that the placements of MR support can better adapt computationally to the complex dynamics of group conversations.

\vspace{2px}\noindent{\bf Risk of excessive agency.}
The MR textual support dynamically rendered by \sysname~may introduce agency risks if users interpret the suggestions as directives to follow rather than information to consider, as reported by some participants.
While this limitation has been discussed in Section~\ref{sec::discussion::implication}, future work may further explore the impact related to the locus of agency; an improved \sysname~ should generate and render support that allows \sysname~to preserve conversation ownership while offering useful assistance. Following this direction, future work may also explore how the level of agency can be adapted to the \sysname~user over extended periods of use.

\vspace{2px}\noindent{\bf Richer nonverbal behavior understanding.}
The current implementation of \sysname~only considers speech emotion and attention formation, with tracked headpose direction serving as a proxy for non-MR users' gaze, which may not precisely correspond to actual gaze direction.
Future work could incorporate a broader range of nonverbal signals from all conversation participants, \eg~hand gestures and f-formations~\cite{Kendon1990, Kendon1990spatial}, enabling the agent to interpret implicit social cues and provide more nuanced conversational support.
Future work may also explore the integration of diverse sensing modalities, \eg~\cite{Boovaraghavan2023Mites, Agarwal2019, Agarwal2020, Chen2020CapTag}, to enhance the contextual awareness of \sysname.

\vspace{2px}\noindent{\bf Scalability to longer conversations.}
Our experimental group conversations last around $15$~min; longer conversations are out of our scope.
It remains unclear how effectively \sysname~can support lengthy conversations lasting several hours or how well the agent can retain key points throughout extended interactions.
Future work could employ more advanced memory mechanisms within AI agents (\eg~\cite{Huang2026, Luo2026}) to maintain coherent, long-term conversational support.

\vspace{2px}\noindent{\bf Study.}
Our study involved six participant groups totaling 18 participants, with findings primarily grounded in qualitative data.
 Future work could include a larger and more diverse participant pool encompassing a wider range of backgrounds (\eg~personality, topic familiarity, and communication skills), as well as evaluate \sysname~with domain-specific discussion topics and collaborative activities involving complicated joint tasks.
\section{Conclusion}\label{sec::conclusion}
\sysname~is a novel MR-based proactive assistive AI agent system for in-person small-group conversation experience.
Our within-subject studies alongside focus groups with six participant groups ($N = 18$) revealed seven key benefits that \sysname~provides in supporting users' engagement and contribution to in-person small-group conversations.
Our findings also highlight directions for future improvements to \sysname~ and for MR support in in-person group conversations.
\begin{acks}

This work was supported by the startup grant provided by \textbf{F}lorida \textbf{I}nternational \textbf{U}niversity (FIU) and the facilities provided by FIU's iCAVE~\cite{icave}. 
We thank the reviewers and our shepherd for their valuable feedback. 
We thank Ariana Taglioretti, Jennifer Large, Theodore Atis, and Steve Luis for their assistance with study logistics. We thank Pedro Remior and Nirmala Arunachalam for their valuable brainstorming support.
We appreciate the early-stage brainstorming contributions of the FIU Capstone II students, \incl~Peace Passos, Ana Morales, Jonathan Marinez, Sabrina Alvarado, and Karen Rivera. 

\end{acks}

\clearpage
\balance
\bibliographystyle{ACM-Reference-Format}
\bibliography{reference}

\newpage
\appendix
\section{Disclaimer}\label{sec::app::ethical_statement}
All studies were approved by the \textbf{I}nstitutional \textbf{R}eview \textbf{B}oard (IRB).
All participants were required to read and sign the approved informed consent before each study session.
Consent was obtained from participants appearing in the demonstrative figures.
Each study session in Section~\ref{sec::formative} and Section~\ref{sec::evaluation} lasted approximately $90$~minutes.
\section{Implementation and Calibration}\label{app::implementation}

This section provides supplementary materials for \S~\ref{sec::system}, detailing the implementation and calibration processes.

\vspace{2px}
\noindent{\bf Implementation.} 
\sysname~ is designed as a PC extended reality system, offloading rendering to a separate GPU-enabled host machine via Meta Quest Link~\cite{QuestLink2025}.
A Unity-based application was developed using the Meta XR All-In-One SDK (version 85.0)~\cite{MetaSDK} to render real-time conversational support information and to control the Orbbec Femto Bolt camera for tracking the poses of non-supported conversation participants.
A Python backend was implemented to transcribe and diarize speech from all conversation participants using Azure's real-time speech and diarization SDK~\cite{AzureSpeechService}, maintain the group conversational context, generate $PT_{role}$ and $PT_{inf}$, and infer proactive decisions and support information using an LLM-based agent.
GPT-4.1~\cite{gpt-4-1} was used during prototyping and study.

\vspace{2px}\noindent{\bf Calibration.}
Calibration was required before each user study session (\S~\ref{sec::evaluation}).
Calibrations of \sysname~follow two steps:

\vspace{2px}\noindent$\bullet$~{\bf Calibration of RGB-D camera.} 
Calibrating the RGB-D camera enables \sysname~to capture the gaze directions of non-MR conversation participants within the same coordinate system as the MR headset.
To facilitate pose tracking, a virtual RGB-D camera is manually positioned so that the tracked positions of the eye and nose of all conversation participants are aligned with their physical counterparts (Figure~\ref{fig::setup}b).
We used one controller as the anchor to calibrate the RGB-D camera, streamlining the \sysname~setup (Figure~\ref{fig::setup}a).

\vspace{2px}\noindent$\bullet$~{\bf Mapping of the conversation participant, diarized speaker ID, and tracked body ID.} 
Our second step of calibration results in a mapping among the participant's preferred name, the speaker ID generated by the Azure diarization pipeline~\cite{AzureSpeechService}, and the body ID tracked by the RGB-D camera.
Before using \sysname, each conversation participant is required to read a designated paragraph aloud, taking turns and including their preferred name.
The \sysname~system then automatically establishes an internal mapping, which is used by the designed agentic pipeline (Figure~\ref{fig::architecture}).

\begin{figure}[h!]
    \centering
    \includegraphics[width=\linewidth]{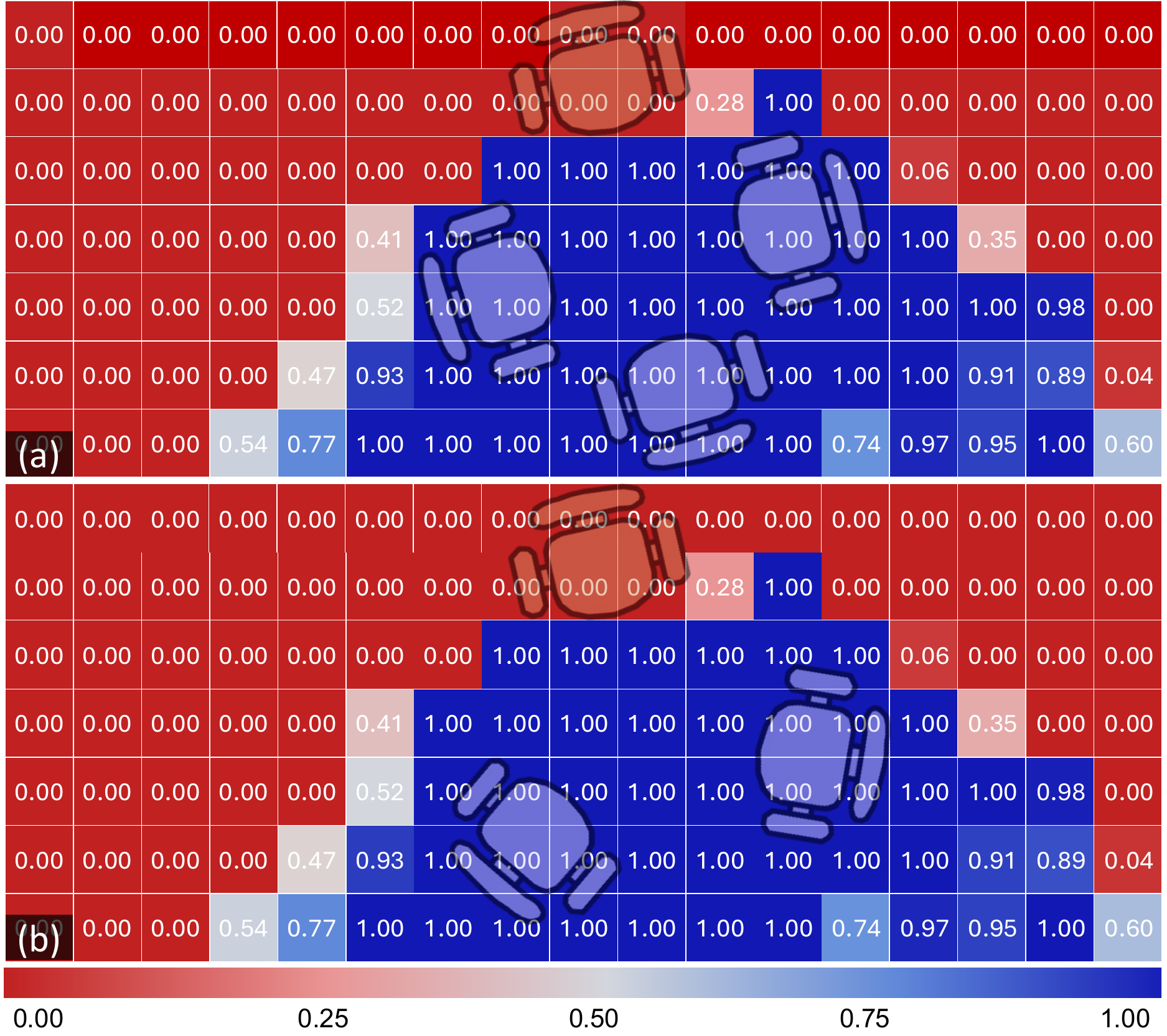}
    \vspace{-0.25in}
    \caption{Tracking performance of the RGB-D camera and the spatial arrangement for (a) four-person and (b) three-person conversations. The top red chair represents the MR user. \red{Numbers in each cell represent the measured tracking confidence score.}}
    \label{fig::app::camera-benchmark}
\end{figure}

{\color{black}

\section{Tracking Performance of the RGB-D Camera}\label{app::track_performance}

This section presents supplementary benchmarking results for the Orbbec camera~\cite{orbbecCam} used to track the head poses of conversation partners.
Our goal is to understand how reliably the instrumented RGB-D camera can track participants across the seating area and to determine the spatial arrangement of conversation participants.
We divided the experimental space into a $7 \times 17$ grid, and, for each cell $(x,y)$, computed a \textbf{tracking confidence score} ($f_{x, y}$).
Since we focused on participants' gaze direction, we computed $f_{x, y}$ using only the key joints ($J$) most relevant to gaze estimation: the head, nose, left eye, and right eye.
The tracking confidence score at frame $i$ ($f_{x,y,i}$) is approximated by Equation~(\ref{eqn::tracking_confidence_per_frame}).
Here, $f_{x,y,i,j}$ denotes the inferred tracking confidence for joint $j$, taking values in $\{0, 1, 2\}$ as provided by Azure Kinect SDK~\cite{azureJointConfidence}.
$f_{x,y,i} = 0$ indicates that all joints in $J$ have a confidence value of $0$ (\ie~ are not tracked) at frame $i$.
$f_{x, y}$ is then computed by Equation~(\ref{eqn::tracking_confidence}), where $N$ is the total number of frames sampled in the one-minute interval, and $f_{\max} = 2$ denotes the maximum joint confidence value.

{\setlength{\abovedisplayskip}{-3pt}
\setlength{\abovedisplayshortskip}{-3pt}
\setlength{\belowdisplayskip}{3pt}
\setlength{\belowdisplayshortskip}{3pt}
\begin{equation}
    f_{x,y,i} = 
        \frac{1}{|J|} \sum_{j \in J} f_{x,y,i,j} 
    \label{eqn::tracking_confidence_per_frame}
\end{equation}

\begin{equation}
    f_{x,y} = \frac{1}{N \times f_{\max}} \sum_{i=1}^N f_{x,y,i}
    \label{eqn::tracking_confidence}
\end{equation}
}

Figure~\ref{fig::app::camera-benchmark} shows the results for the four-person and three-person conversation layouts, along with the corresponding seat arrangements.
The MR user's seat was intentionally placed outside the camera's field of view, whereas the non-MR users were placed in areas with high tracking confidence.
}

\section{Interpersonal Communication Skills Inventory~(ICSI) Questionnaires}\label{app::questionaires}
The \textbf{I}nterpersonal \textbf{C}ommunication \textbf{S}kills \textbf{I}nventory~(ICSI)~\cite{Interpersonal-Communication-Skills-Inventory, Bienvenu1971, Boyd1977} was used in studies described in \S~\ref{sec::formative} and \S~\ref{sec::evaluation}.
This standardized questionnaire was used to assess participants' communication skills, which is structured into four sections: \emph{sending clear messages}, \emph{listening}, \emph{giving and receiving feedback}, and \emph{handling emotional interactions}.
All questions were rated on a scale of \emph{usually}, \emph{sometimes}, and \emph{seldom}. Table~\ref{tab::icsi_scoring_key} shows the scoring key.

\vspace{4px}\noindent\textbf{Section I: Sending Clear Messages}

\begin{itemize}[leftmargin=*]
    \item \textbf{Q1.} Is it difficult for you to talk to other people?
    
    \item \textbf{Q2.} When you are trying to explain something, do others tend to put words in your mouth or finish your sentences for you?
    
    \item \textbf{Q3.} In conversation, do your words usually come out the way you would like?
    
    \item \textbf{Q4.} Do you find it difficult to express your ideas when they differ from the ideas of people around you?
    
    \item \textbf{Q5.} Do you assume that the other person knows what you are trying to say, and leave it to him/her to ask you questions?
    
    \item \textbf{Q6.} Do others seem interested and attentive when you are talking to them?
    
    \item \textbf{Q7.} When speaking, is it easy for you to recognize how others are reacting to what you are saying?
    
    \item \textbf{Q8.} Do you ask the other person to tell you how she/he feels about the point you are trying to make?
    
    \item \textbf{Q9.} Are you aware of how your tone of voice may affect others?
    
    \item \textbf{Q10.} In conversation, do you look to talk about things of interest to both you and the other person?
\end{itemize}

\vspace{4px}\noindent\textbf{Section II: Listening}

\begin{itemize}[leftmargin=*]

    \item \textbf{Q11.} In conversation, do you tend to do more talking than the other person does?

    \item \textbf{Q12.} In conversation, do you ask the other person questions when you don't understand what they've said?

    \item \textbf{Q13.} In conversation, do you often try to figure out what the other person is going to say before they've finished talking?

    \item \textbf{Q14.} Do you find yourself not paying attention while in conversation with others?

    \item \textbf{Q15.} In conversation, can you easily tell the difference between what the person is saying and how he/she may be feeling?

    \item \textbf{Q16.} After the other person is done speaking, do you clarify what you heard them say before you offer a response?

    \item \textbf{Q17.} In conversation, do you tend to finish sentences or supply words for the other person?

    \item \textbf{Q18.} In conversation, do you find yourself paying most attention to facts and details, and frequently missing the emotional tone of the speakers' voice?

    \item \textbf{Q19.} In conversation, do you let the other person finish talking before reacting to what she/he says?

    \item \textbf{Q20.} Is it difficult for you to see things from the other person's point of view?

\end{itemize}

\vspace{4px}\noindent\textbf{Section III: Giving and Getting Feedback}

\begin{itemize}[leftmargin=*]

    \item \textbf{Q21.} Is it difficult to hear or accept constructive criticism from the other person?

    \item \textbf{Q22.} Do you refrain from saying something that you think will upset someone or make matters worse?

    \item \textbf{Q23.} When someone hurts your feelings, do you discuss this with him/her?

    \item \textbf{Q24.} In conversation, do you try to put yourself in the other person's shoes?

    \item \textbf{Q25.} Do you become uneasy when someone pays you a compliment?

    \item \textbf{Q26.} Do you find it difficult to disagree with others because you are afraid they will get angry?

    \item \textbf{Q27.} Do you find it difficult to compliment or praise others?

    \item \textbf{Q28.} Do others remark that you always seem to think you are right?

    \item \textbf{Q29.} Do you find that others seem to get defensive when you disagree with their point of view?

    \item \textbf{Q30.} Do you help others to understand you by saying how you feel?

\end{itemize}

\vspace{4px}\noindent\textbf{Section IV: Handling Emotional Interactions}

\begin{itemize}[leftmargin=*]

    \item \textbf{Q31.} Do you have a tendency to change the subject when the other person's feelings enter into the discussion?

    \item \textbf{Q32.} Does it upset you a great deal when someone disagrees with you?

    \item \textbf{Q33.} Do you find it difficult to think clearly when you are angry with someone?

    \item \textbf{Q34.} When a problem arises between you and another person, can you discuss it without getting angry?

    \item \textbf{Q35.} Are you satisfied with the way you handle differences with others?

    \item \textbf{Q36.} Do you sulk for a long time when someone upsets you?

    \item \textbf{Q37.} Do you apologize to someone whose feelings you may have hurt?

    \item \textbf{Q38.} Do you admit that you're wrong when you know that you are/were wrong about something?

    \item \textbf{Q39.} Do you avoid or change the topic if someone is expressing his or her feelings in a conversation?

    \item \textbf{Q40.} When someone becomes upset, do you find it difficult to continue the conversation?

\end{itemize}

\begin{table*}
    \centering
    \caption{Assigned scores for responses on the Interpersonal Communication Skills Inventory~\cite{Interpersonal-Communication-Skills-Inventory, Bienvenu1971, Boyd1977}.}
    \vspace{-.1in}
    \includegraphics[width=\linewidth]{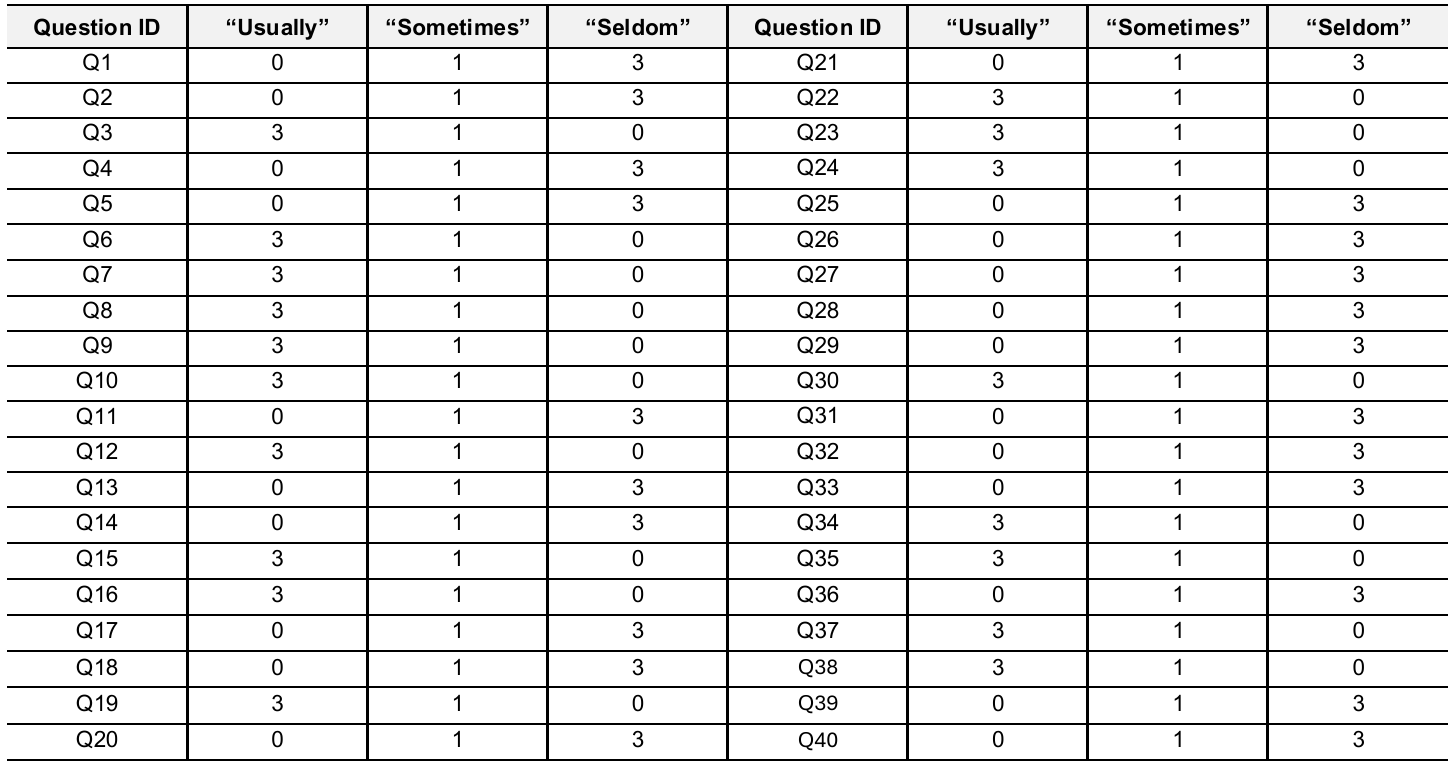}
    \label{tab::icsi_scoring_key}
\end{table*}

\section{Participant Demographics}\label{app::participants}

This section describes the demographics of the recruited participants and the composition of participant groups. Participants were assigned to groups based on their scheduling availability.
We use FG\# and FP\# to index the participant group and participant in the formative study (\S~\ref{sec::formative}).
Likewise, G\# and P\# are used to index participant group and participant in the final user study (\S~\ref{sec::evaluation}).
Table~\ref{tab::formative-participants} shows the demographics of the recruited participants in the formative study (\S~\ref{sec::formative}).
Participants using PT1 and PT2 are highlighted in \colorbox{mr-tp1}{green} and \colorbox{mr-tp2}{blue}, respectively.
Table~\ref{tab::participant} shows the demographics of the recruited participants involved in the user study described in \S~\ref{sec::evaluation}.
Participants wearing the MR headset are highlighted in \colorbox{mr}{orange}.
P18 withdrew from the study due to an unexpected event. As a result, one of the researchers served as a non-MR conversation participant, but did not participate in or influence the post-study focus group discussion.

\begin{table*}[t]
    \centering
    \caption{Demographics of the formative study participants. Participants using PT1 and PT2 are highlighted in \colorbox{mr-tp1}{green} and \colorbox{mr-tp2}{blue}, respectively. Responses of \emph{AI Experience} and \emph{Blind Typing} reflect participants' agreement with the statements \emph{``I am experienced in using AI tools''} and \emph{``I am experienced in typing without looking at the keyboard,''} respectively, with $\bm{5}$ indicating \emph{strongly agree} and $\bm{1}$ indicating \emph{strongly disagree}. Each Interpersonal Communication Skill Inventory~\cite{Interpersonal-Communication-Skills-Inventory, Bienvenu1971, Boyd1977} score ranges from $\bm{0}$ to $\bm{30}$, with higher scores indicating areas of strength. }
    \vspace{-0.1in}
    \includegraphics[width=\linewidth]{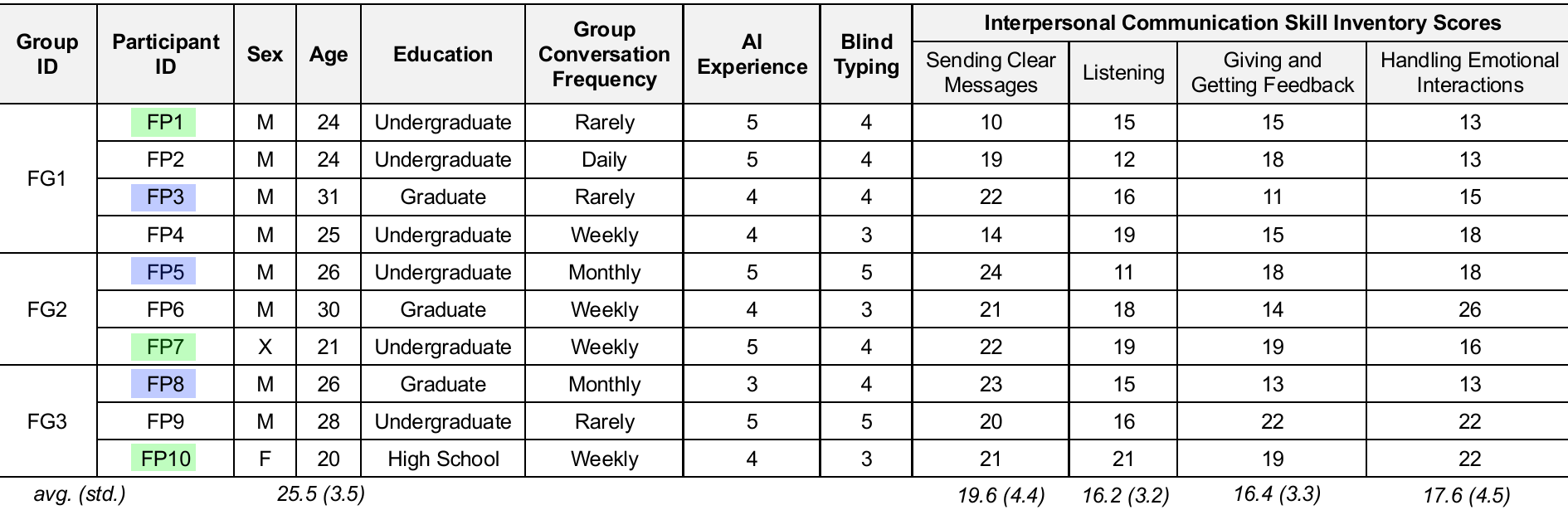}
    \label{tab::formative-participants}
\end{table*}

\begin{table*}[t]
    \centering
    \caption{Demographics of the participants for evaluation. Participants wearing the MR headset are highlighted with \colorbox{mr}{orange}. Responses of \emph{AI experience} and \emph{topic familiarity} reflect participants' agreement with the statements \emph{``I am experienced in using AI tools''} and \emph{``I am familiar with the given topic,''} respectively, with $\bm{5}$ indicating \emph{strongly agree} and $\bm{1}$ indicating \emph{strongly disagree}. Each Interpersonal Communication Skill Inventory~\cite{Interpersonal-Communication-Skills-Inventory, Bienvenu1971, Boyd1977} score ranges from $\bm{0}$ to $\bm{30}$, with higher scores indicating areas of strength. P18 withdrew from the study due to an unexpected event; one of the researchers served as the unsupported conversation participant, but did not participate in or influence the post-study focus group discussion.}
    \vspace{-0.1in}
    \includegraphics[width=\linewidth]{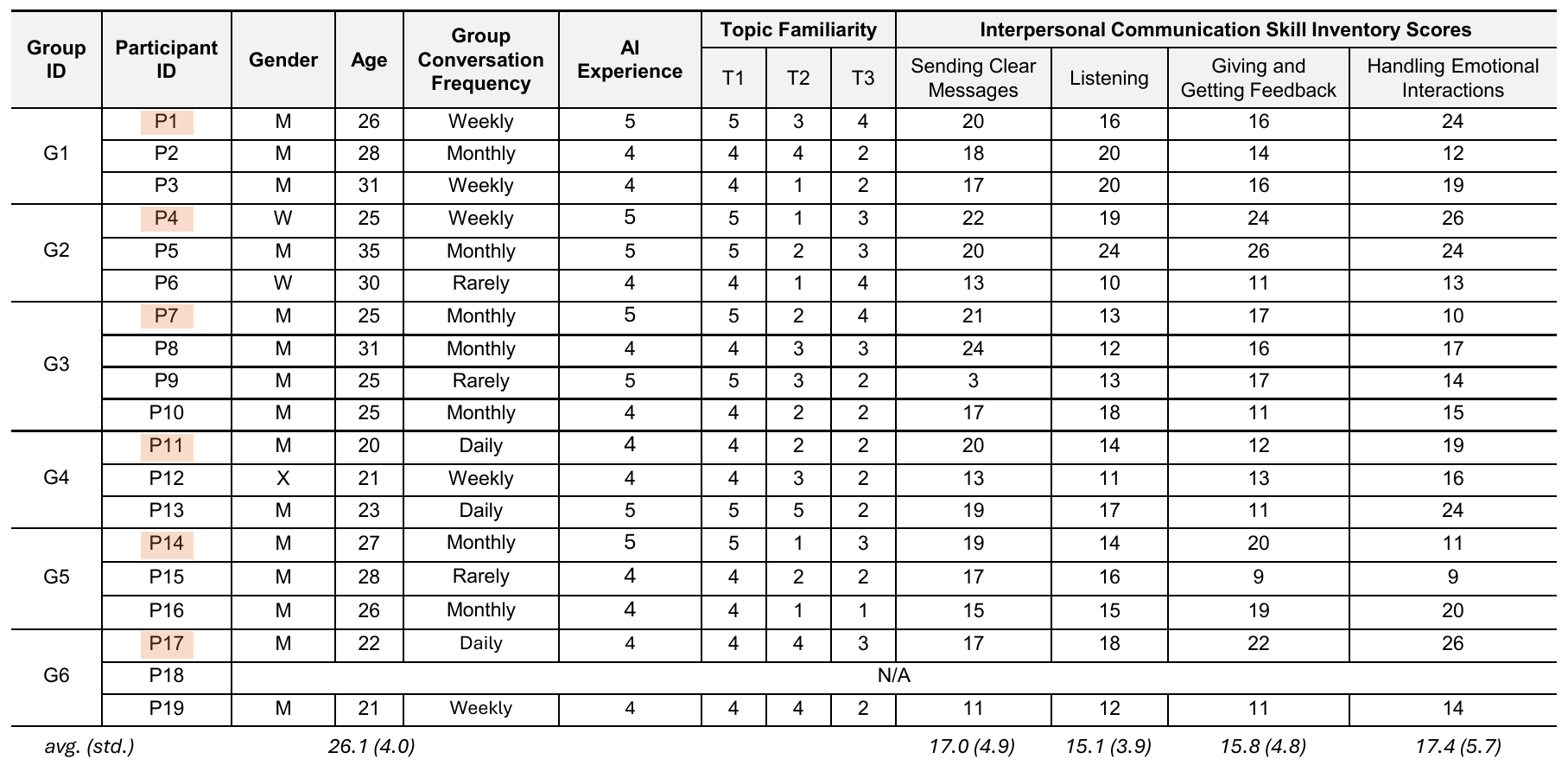}
    \label{tab::participant}
\end{table*}
\section{Survey Responses}\label{app::survey}

\begin{figure*}
    \centering
    \includegraphics[width=\linewidth]{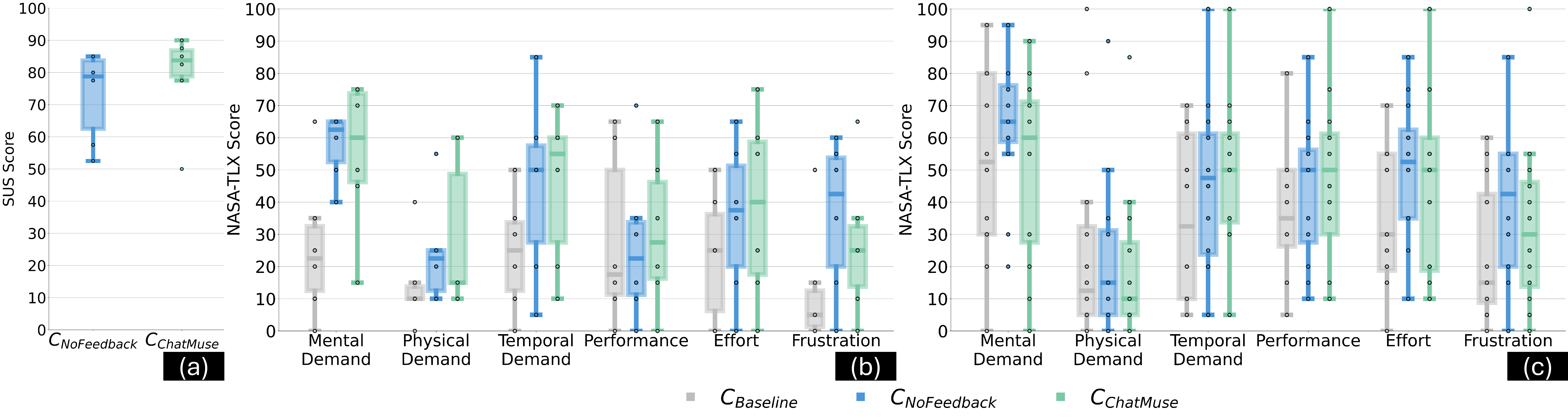}
    \vspace{-0.2in}
    \caption{Survey responses of SUS and NASA TLX. (a) Responses of SUS from MR users. (b) Responses of NASA TLX from participants wearing MR headset. (c) Responses of NASA TLX from participants not wearing MR headset. }
    \label{fig::nasa-tlx-sus}
\end{figure*}

This section describes participants' post-session survey responses.
Figure~\ref{fig::nasa-tlx-sus}a shows the responses of \textbf{S}ystem \textbf{U}sability \textbf{S}cale (SUS) reported by MR users for $C_{NoFeedback}$ and $C_{\sysname}$.
\mbox{Figure~\ref{fig::nasa-tlx-sus}b--c} describe the responses of \textbf{NASA} \textbf{T}ask \textbf{L}oad Inde\textbf{x}~(NASA TLX) from both MR user and participants not wearing the headset.
No statistically significant differences ($p > .05$) were found across any of the six NASA TLX dimensions for participants who did not wear the MR headset (Figure~\ref{fig::nasa-tlx-sus}c).

\end{document}